\documentclass[]{spie}  

\usepackage{amsmath,amsfonts,amssymb}
\usepackage{graphicx}
\usepackage{multirow}
\usepackage{float}
\usepackage{subcaption}
\usepackage{lineno}
\usepackage{booktabs}
\usepackage{tabularx}

\usepackage[colorlinks=true, allcolors=blue]{hyperref}

\title{The Double-Sided Silicon Strip Detector Tracker onboard the  
ComPair Balloon Flight}

\author[a,b]{Nicholas Kirschner}
\author[b]{Carolyn Kierans}
\author[b,d,f]{Sambid Wasti}
\author[b,g]{Adam J. Schoenwald}
\author[b]{Regina Caputo}
\author[e]{Sean Griffin}
\author[b,f]{Iker Liceaga-Indart}
\author[h]{Lucas Parker}
\author[b]{Jeremy S. Perkins}
\author[b,c,d]{Anna Zajczyk}

\affil[a]{The Department of Physics, The George Washington University, 725 21st St NW, Washington, DC 20052, USA}
\affil[b]{NASA Goddard Space Flight Center, Greenbelt, MD, USA}
\affil[c]{Center for Space Sciences and Technology, University of Maryland, Baltimore County, 1000 Hilltop Circle, Baltimore, MD 21250, USA}
\affil[d]{Center for Research and Exploration in Space Science and Technology, NASA/GSFC, Green- belt, MD 20771, USA}
\affil[e]{Wisconsin IceCube Particle Astrophysics Center, University of Wisconsin-Madison, 222 W Washington Ave Unit 500, Madison, WI 53703, USA}
\affil[f]{Catholic University of America, 620 Michigan Ave NE, Washington, DC 20064, USA}
\affil[g]{University of Maryland, Baltimore County, 1000 Hilltop Circle, Baltimore, MD 21250, USA}
\affil[h]{Los Alamos National Laboratory, Los Alamos, NM 87544, USA}

\authorinfo{Further author information: (Send correspondence to Nicholas J Kirschner)\\N.J.K.: E-mail: nkirschner@gwu.edu, Telephone: 1 215 290 3268}

\begin{document} 
\maketitle

\begin{abstract}
\begin{sloppypar}
{The \textit{ComPair} balloon instrument is a prototype of the All-sky Medium Energy Gamma-ray Observatory (AMEGO) mission concept. AMEGO aims to bridge the spectral gap in sensitivity that currently exists from $\sim$100 keV to $\sim$100 MeV by being sensitive to both Compton and pair-production events. This is made possible through the use of four subsystems working together to reconstruct events: a double-sided silicon strip detector (DSSD) Tracker, a virtual Frisch grid cadmium zinc telluride (CZT) Low Energy Calorimeter, a ceasium iodide (CsI) High Energy Calorimeter, and an anti-coincidence detector (ACD) to reject charged particle backgrounds. Composed of 10 layers of DSSDs, \textit{ComPair's} Tracker is designed to measure the position of  photons that Compton scatter in the silicon, as well as reconstruct the tracks of electrons and positrons from pair-production as they propagate through the detector. By using these positions, as well as the absorbed energies in the Tracker and 2 Calorimeters, the energy and direction of the incident photon can be determined. This proceeding will present the development, testing, and calibration of the \textit{ComPair} DSSD Tracker and early results from its balloon flight in August 2023.}
\end{sloppypar}
\end{abstract}


\section{INTRODUCTION}
\label{sec:intro}  

Since the discovery of gamma-ray astrophysical sources such as pulsars, active galactic nuclei (AGN), and gamma-ray bursts (GRBs) in the mid-20th century, the field of gamma-ray astrophysics has offered scientists the promising ability to learn more about our ever-expanding universe. While gamma-ray science has come a long way over the past several decades, current instrument limitations leaves plenty of questions unanswered such as: what are the underlying emission mechanisms behind GRB jets, or what is the composition of the diffuse MeV background?\cite{diwan2023} To better answer these questions, better coverage of the MeV energy regime is necessary. To date, the gamma-ray sky from a couple hundred keV to a hundred MeV has been vastly unexplored due to competing photon-interaction processes. While Compton scattering dominates $\sim$1 MeV, and electron-positron pair-production dominates $\sim$100 MeV, there's a cross-over in between where the two compete. Therefore, in order to have an instrument sensitive in this energy range, it must be optimized for both Compton and pair-production events.

\subsection{AMEGO: Bridging the Gap in Data}

\begin{figure}[H]
  \centering
  \begin{minipage}[b]{0.49\textwidth}
    \includegraphics[width=\textwidth]{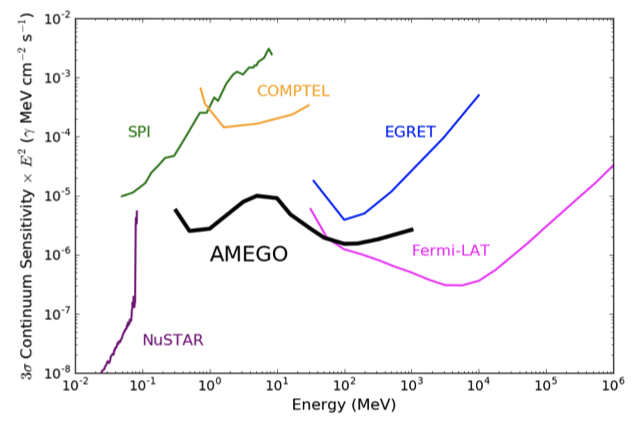}
    \subcaption{Simulated sensitivity of AMEGO}
    \label{fig:AMEGO_coverage}
  \end{minipage}
  \hfill
  \begin{minipage}[b]{0.49\textwidth}
    \includegraphics[width=0.8\textwidth]{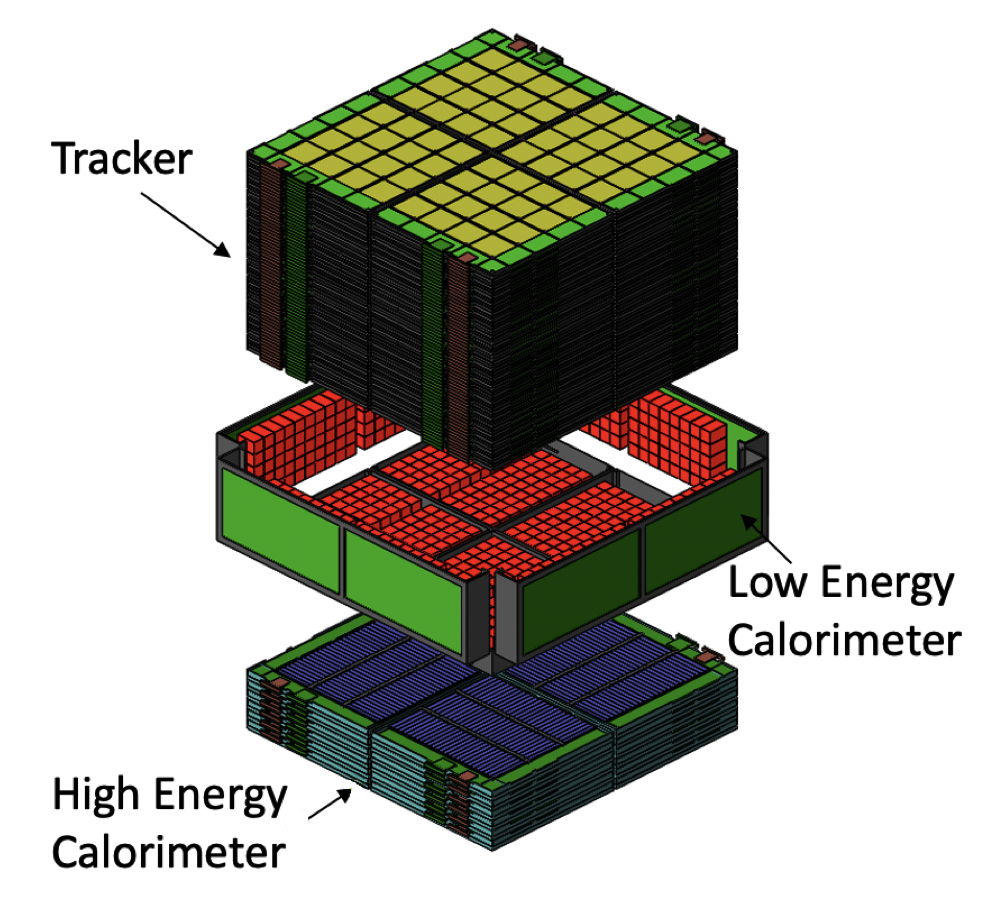}
    \subcaption{Design for AMEGO’s detector}
    \label{fig:AMEGO_Layout}
  \end{minipage}
  \caption{(a) AMEGO will provide coverage of the MeV band with sensitivity at least 20 times better than {\it COMPTEL}\cite{mcenery2019allsky}. (b) AMEGO is comprised of four subsystems: a DSSD Tracker, a Low Energy CZT Calorimeter, a High Energy CsI calorimeter, and an ACD (not pictured) that encapsulates the other subsystems.}
\end{figure}

The All-sky Medium Energy Gamma-ray Observatory (AMEGO) is a gamma-ray mission concept\cite{Kierans_2020, mcenery2019allsky} that aims to cover the infamous MeV gap with sensitivity over an order of magnitude better than previous instruments (Fig.~\ref{fig:AMEGO_coverage}). AMEGO will be able to resolve both Compton and pair-production events due to its detector design which is comprised of four subsystems (Fig.~\ref{fig:AMEGO_Layout}): a double-sided silicon-strip detector (DSSD) Tracker, a virtual Frisch grid cadmium zinc telluride (CZT) Low Energy Calorimeter, a caesium iodide (CsI) High Energy Calorimeter, and an anti-coincidence detector (ACD).

\subsection{The \textit{ComPair} Prototype}

\begin{figure}[H]
  \centering
  \begin{minipage}[b]{0.57\textwidth}
    \includegraphics[width=\textwidth]{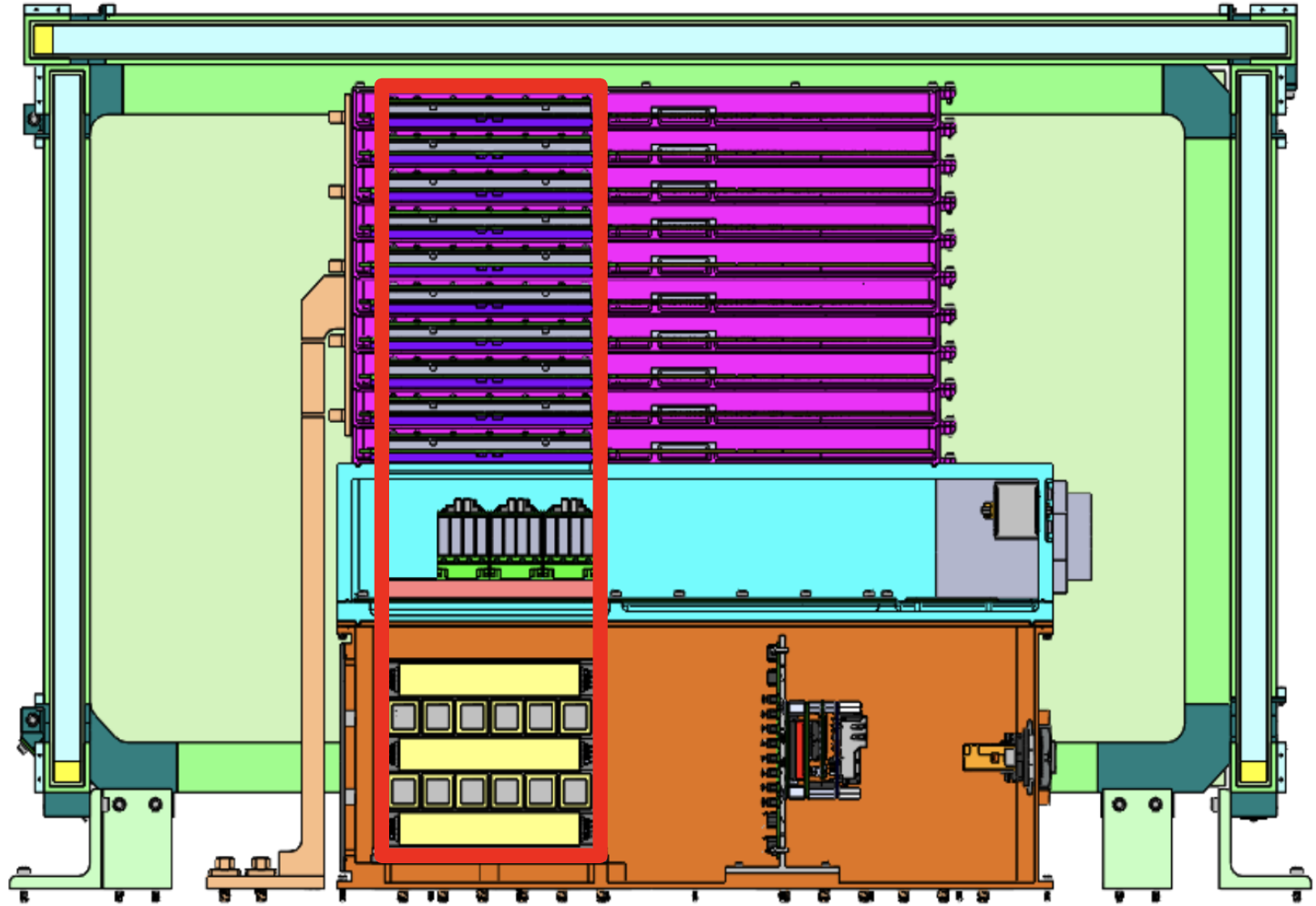}
    \subcaption{CAD model of {\it ComPair}}
    \label{fig:ComPair_CAD}
  \end{minipage}
  \hfill
  \begin{minipage}[b]{0.42\textwidth}
    \includegraphics[width=50mm]{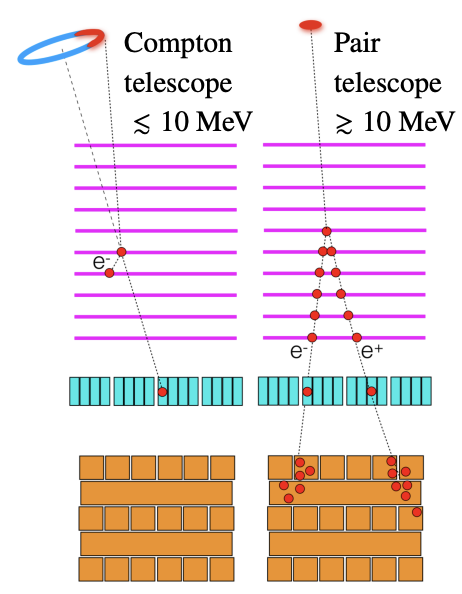}
    \subcaption{Photon interactions in {\it ComPair}'s detectors}
    \label{fig:compton-pair}
  \end{minipage}
  \caption{(a) CAD model of {\it ComPair} with the active area boxed in red. Each subsystem's detectors and electronics are integrated into aluminum enclosures stacked on top of each other, including the DSSD Tracker (magenta), the CZT Calorimeter (teal), and the CsI Calorimeter (orange). Surrounding the instrument stack is the ACD (green) that rejects charged particles. Attached to the side of the Tracker are thermal straps (copper) to dissipate heat. (b) {\it ComPair}'s design enables reconstruction of both Compton and pair events. Below $\sim$10 MeV, photons are more likely to Compton scatter in the Tracker (shown in magenta), before the scattered photon gets absorbed in the CZT Calorimeter (in teal). Above $\sim$10 MeV, photons will likely convert into an electron-positron pair in the Tracker, and the pair products are tracked through the CZT Calorimeter before being absorbed in the CsI Calorimeter (in orange). \cite{Valverde_2023}}
\end{figure}

\textit{ComPair} was funded in 2015 to develop a prototype of the AMEGO telescope and to test the detector system in a space-like environment through a high-altitude balloon flight. \textit{ComPair} is composed of smaller versions of AMEGO's four subsystems (Fig.~\ref{fig:ComPair_CAD}). The DSSD Tracker\cite{Griffin:ComPair} is comprised of 10 single-detector layers, stacked on top of each other. The CZT calorimeter \cite{FrischGridCZT}, composed of 9 modules of 4$\times$4 arrays of CZT bars, is placed underneath the Tracker. The CsI calorimeter, sitting below the Tracker and CZT calorimeter, consists of 5 layers of six CsI bars where each layer is orthogonal to its neighboring layers in a hodoscopic arrangement \cite{Woolf_2018, Shy_2023}. The {\it ComPair} design is similar to previous instruments that were  also designed to cover the Compton and pair production regime: MEGA and TIGRE \cite{Bloser_2003, TIGRE}.
 
Below 10 MeV, $\gamma$-ray photons will predominantly Compton scatter in the detector \cite{Kierans_2022}. Above 10 MeV, $\gamma$-rays will often interact with the detector to produce pairs \cite{Thompson_2022}, therefore, to be sensitive in the MeV regime, {\it ComPair} must measure both Compton and pair-production events. For Compton events, a photon interacts with at least one of the Tracker layers, scattering an electron in the process, before being absorbed by the CZT and/or CsI Calorimeter (Fig.~\ref{fig:compton-pair}). The energy of the initial photon is then found by summing the energies of each interaction. The positions of the photon interactions are used to recover the initial direction of the incident photon via the Compton equation:
\begin{equation}
\label{eq:Compton}
E' = \frac{E}{1 + \frac{E}{mc^2}(1 - \cos{\theta})}
\end{equation}

\noindent where $E'$ is the energy of the photon after scattering, $E$ is the initial photon energy, and $\theta$ is the scattering angle. The maximum energy that can be transferred in a Compton interaction occurs when $\theta$ = 180\textdegree and the photon back-scatters. For pair-production events, a photon interacts with a Tracker layer resulting in an electron and positron pair. This pair then gets tracked through the Tracker, before being absorbed in the two Calorimeters. Once again, the incident photon energy is then found simply by summing the energies of each interaction. The trajectory of both the positron and electron is then used to recover the initial direction via conservation of momentum (Fig.~\ref{fig:compton-pair}).

\section{THE SILICON TRACKER}

The Tracker provides the most precise position measurements of any of the subsystems, which is vital in reconstructing the original photon direction. The following sections will detail the DSSD design, the accompanying front-end electronics, and the integration and testing of the {\it ComPair} Tracker.

As mentioned above, the Tracker needs to act as both Compton scattering and pair creation material, that can achieve good energy and position resolution. To measure energies adequately, a semiconductor detector is necessary. Sufficient position resolution necessitates the need for fine segmentation in the detector. The need to track the trajectory of charged particles (such as electron-positron pairs) as well as to track Compton events is achieved with many two-dimensional segmented layers. With these requirements in mind, double-sided silicon strip detectors were selected. Additionally, silicon offers a large cross section for Compton events and has been used before in spaceflight instrumentation. 

\subsection{Double-sided Silicon Strip Detectors}

The {\it ComPair} Tracker uses Micron Semiconductor TTT13 double-sided silicon strip detectors. Each detector is $10 \times 10$~cm$^{2}$ and $500 ~\mu\text{m}$ thick with 192 orthogonally-oriented strips per side, with a pitch of $510 ~\mu\text{m}$.  The orthogonal strips allow for the measurement of the X and Y position of interaction within the wafer. 

The n-type bulk silicon wafers have high resistivity, generally $> 20$~k$\Omega$-cm.  The ohmic side of the detector (referred to as n-side) has n-type strips isolated with a p-type implant, either through the p-stop or p-spray technique. The detectors are reverse biased with a positive voltage (+60 V) applied to the ohmic side to fully deplete the  $500 ~\mu\text{m}$ depth of the detector, and isolate the n-side strips. The strips on both sides are AC coupled through a polysilicon resistor and coupling capacitor, with DC pads accessible for probe-station testing.

\begin{figure}[H]
\centering
    \includegraphics[width=120mm]{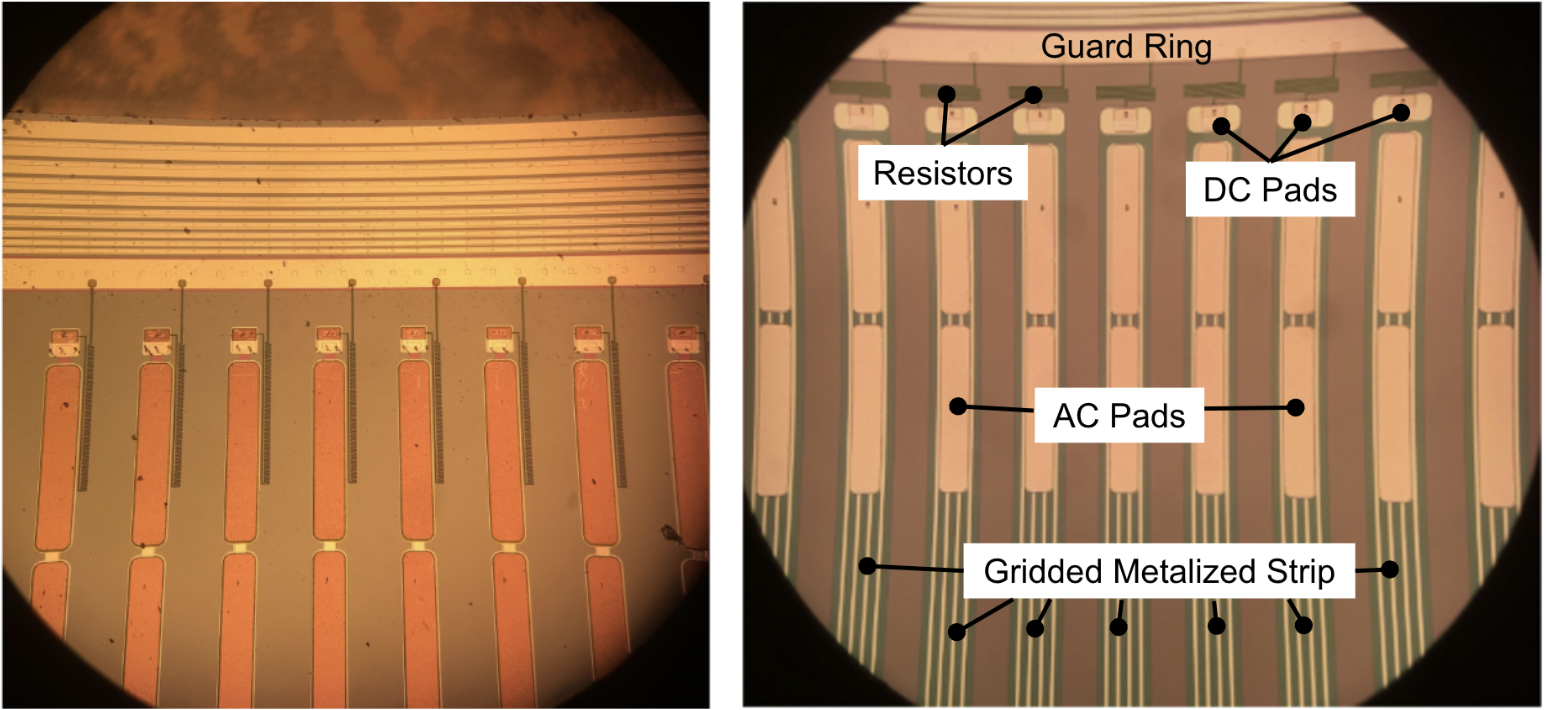}
\caption{(Left) One of the original designs of the Tracker detector. This shows the n-side of the detector with a 60~$\mu$m strip pitch and p-stop isolation. (Right) The labelled final design of the Tracker detectors with a gridded metalization layer on 310~$\mu$m strips with p-spray isolation. An assortment of detector designs were used in the 10 layer Tracker, with 7/10 being this final design and 3/10 being the previous design.}
\label{fig:detector_image}
\end{figure}

The {\it ComPair} DSSD design went through multiple revisions with Micron Semiconductor, where the strip width, implant type, and metalization design was modified in an attempt to optimize the performance of the detectors. The key performance consideration is the noise level in these detectors, as a precise measure of the interaction energy is needed for accurate Compton reconstruction. The dominant sources of noise are the shot noise dominated by the leakage current, the thermal noise defined by the polysilicon resistor, and the electronic noise given by the interstrip capacitance. The requirements for these detectors were: leakage current $<$10 nA per strip, bias resistor value $>$30~$\text{M}\Omega$, coupling capacitor 500-1000~pF, and interstrip capacitance $<$10~pF. Microscope images of two different designs for the n-side isolation and strip metalization are shown in Figure~\ref{fig:detector_image}, where the image on the left shows one of the first designs with 60~$\mu$m strip pitch and p-stop isolation, and the image on the right shows a gridded metalization layer on 310~$\mu$m strips with p-spray isolation. 

Upon delivery of the DSSDs from Micron, they were bulk characterized at NASA Goddard Space Flight Center (GSFC) and select wafers underwent detailed single-strip measurements to verify the relevant design parameters and compare with values provided by Micron. A summary of the 10 detectors that were used for the {\it ComPair} tracker are shown in Table~\ref{tab:detectors}. All of the detectors achieved full depletion at or below 55 V, and therefore all could be biased with a single 60 V supply. While the values for the polysilicon resistance and the coupling capacitance (not shown, but on average was $\sim$1000 pF) were within the design specifications, the single-strip leakage current requirement of $<$10~nA was not achieved in any of the detectors. The interstrip capacitance was only measured for a few strips on a handful of detectors; for the 60~$\mu$m strip width the interstrip capacitance was $\sim$4~pF on the junction side and $\sim$12~pF on the ohmic side, while the 310~$\mu$m strips with p-spray isolation showed $\sim$7~pF and $\sim$12~pF on the junction and ohmic side, respectively.

\begin{table}[htbp]
\caption{Description of the 10 detectors used for {\it ComPair} and their properties. The two values in the 7th and 8th columns correspond to the n-side and p-side of the detectors respectively.}
\centering
\resizebox{\textwidth}{!}{%
\begin{tabular}{|c|c|c|c|c|c|c|c|c|}
\toprule
\hline
\multicolumn{1}{|p{1.2cm}}{Layer number} &
\multicolumn{1}{|p{1.4cm}}{TTT13 Wafer \#} &
\multicolumn{1}{|p{1.5cm}}{Isolation technique} &
\multicolumn{1}{|p{1.8cm}}{Strip width (\textmu m)} &
\multicolumn{1}{|p{1.8cm}}{Depletion voltage (V)} &
\multicolumn{1}{|p{2.1cm}}{Bulk Current at 60 V (\textmu A)} &
\multicolumn{1}{|p{3.0cm}}{Average polysilicon resistance (MOhm)} &
\multicolumn{1}{|p{3.1cm}}{Average single-strip leakage current (nA)} &
\multicolumn{1}{|p{1.5cm}|}{\# of strip faults} \\
\midrule
\hline
0 & 3340-24 & P-spray & 60 & 55 & 2.8 & 41 / 33 & 36 / 60 & 15 \\
\hline
1 & 3479-2 & P-spray & 310 & 30 & 2.1 & 55 / 40 & 22 / 27 & 4 \\
\hline
2 & 3479-18 & P-spray & 310 & 40 & 2.3 & 39 / 32 & 25 / 25 & 2 \\
\hline
3 & 3479-6 & P-spray & 310 & 30 & 1.8 & 40 / 33 & 20 / 26 & 9 \\
\hline
4 & 3340-7 & P-stop & 60 & 45 & 3.5 & 45 / 21 & 34 / 22 & 13 \\
\hline
5 & 3392-11 & P-stop & 60 & 55 & 4.7 & 36 / 24 & 32 / 34 & 13 \\
\hline
6 & 3479-11 & P-spray & 310 & 35 & 1.8 & 35 / 33 & 19 / 16 & 2 \\
\hline
7 & 3479-21 & P-spray & 310 & 30 & 1.6 & 39 / 34 & 21 / 25 & 4 \\
\hline
8 & 3479-19 & P-spray & 310 & 30 & 1.6 & 17 / 16 & 22 / 19 & 5 \\
\hline
9 & 3479-10 & P-spray & 310 & 30 & 2.1 & 32 / 32 & 21 / 25 & 7 \\
\hline
\bottomrule
\end{tabular}%
}
\label{tab:detectors}
\end{table}

\begin{figure}[H]
\centering
\vspace{-.05cm}
    \includegraphics[width=100mm]{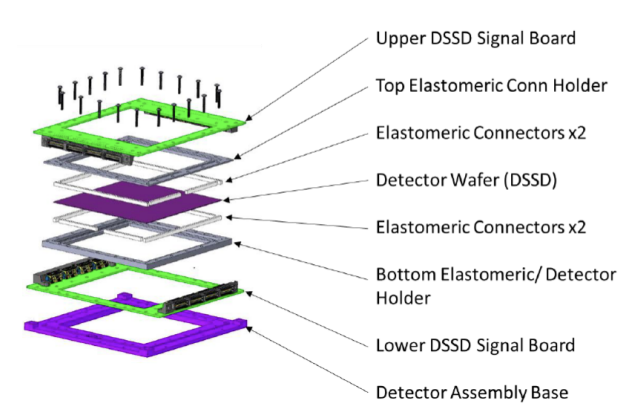}
\caption{An exploded view of the DSSD carrier. The detector assembly base is made out of aluminum and the elastomeric connector holders are made out of delrin. Elastomeric connectors span across the AC-coupled strips on either side of the detector to connect the strips with the upper and lower DSSD signal boards.}
\label{fig:Detector_carrier}
\end{figure}

The detectors are mounted in a custom carrier board using elastomeric electrical contacts to supply the bias voltage and to make a connection with the AC pads on each strip. An expanded view of the carrier board is shown in Fig.~\ref{fig:Detector_carrier}. Strips with faulty coupling capacitors (i.e. a short between DC and AC pads) had to be masked prior to connection to the front end electronics. This was done by applying small pieces of kapton tape to the carrier board to physically isolate the channel, saving the ASIC inputs from the +60 V bias voltage. 

\subsection{Front End Electronics}

\begin{figure}[H]
  \centering
    \includegraphics[width=0.5\textwidth]{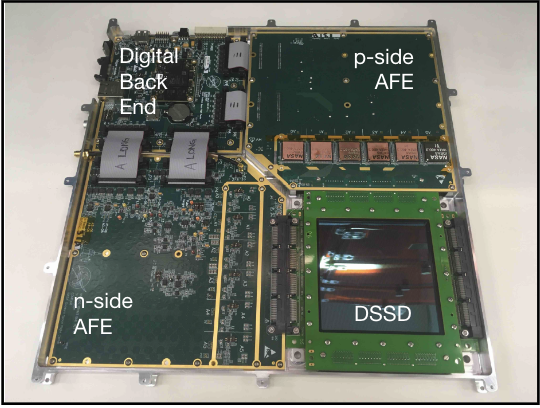}
  \caption{One of {\it ComPair}'s tracking layers detailing the 4 main components. The p-side and n-side AFE boards are identical and can be easily configured for opposite polarity read out.}
    \label{fig:ComPair_Layer}
\end{figure}

Each layer of \textit{ComPair}'s Tracker consists of a DSSD, which sits in a carrier tray along side two identical analog front end (AFE) boards, and a digital back end (DBE) board (Fig.~\ref{fig:ComPair_Layer}) \cite{Kierans_2020}. Ionizing radiation results in a charge induced on the AC-coupled strips, sending a pulse from the DSSD to the AFE board. Each strip has a configurable threshold and comes with the ability to mask (disable) the full channel. Six Application Specific Integrated Circuits (ASICs; IDEAS VATA 460.3 \cite{IDEAS}) are integrated in an AFE board to read out and digitize the pulse height, which is proportional to energy. Each ASIC has 32 inputs with charge-sensitive preamplifiers (CSAs) and contains 32 10-bit ADCs (one per channel, corresponding to the 32 strips) \cite{asic_manual}. Each CSA is connected to a 2 $\mu$s slow shaper for spectroscopy and a 0.6 $\mu$s fast shaper for triggering the readout. The slow shaper is connected to a sample and hold circuit that holds the pulse height when an external hold signal is received. Additionally, each ASIC supports both positive and negative polarity input charges of up to +50/-90 fC, and allows for external calibration charges, subsequently referred to as ``calibration pulses", to be injected for easy functionality testing. Both the DSSDs and the ASICs come from out-of-house suppliers and are thoroughly tested before integration \cite{Griffin:ComPair}. 

The information read out by the ASICs is subsequently transmitted to a field programmable gate array (FPGA; model: Xilinx Zynq 7020) for processing. The DBE board contains the FPGA, responsible for handling the data from all 12 ASICs, as well as telemetry and communication with the data acquisition system and Trigger Module \cite{Makoto_2020}, a peripheral system that checks coincidence of events between subsystems and between Tracker layers. The layers are separated by a distance of 1.9cm (the thickness of a single layer tray) and each layer requires 9.5 W of power.

\subsection{Integration and Testing}

The Tracker layer integration process included integrating the silicon detectors into their respective carriers, followed by the assembly of carriers and electronics into aluminum tray holders. First, the DSSD is placed into the carrier with elastomeric spanning across the AC-coupled strips. Next, the carrier is integrated into the aluminum tray along with the two AFE boards and DBE board. Once complete, the layer is ready for testing.

For testing each of the Tracker layers, a calibration charge is injected into each silicon strip to ensure that the ASICs read out each strip correctly before moving on to the masking stage. In the masking stage, each strip in each DSSD is examined, and its triggering threshold is raised until its triggering rate in the absence of a gamma-ray source is $\lesssim$3 Hz. Due to the high leakage current in strips and additional fabrication issues, some strips have noise levels so high that the strip is inoperable. Since these strips are constantly triggering on noise, their corresponding ASIC channels can be turned off, or ``masked". Due to instabilities in the detector, it was found that the baseline noise for certain strips changes over time. It is for this reason that the entire process of monitoring strip trigger rates, raising thresholds, and masking the overly noisy channels was ongoing up until launch.

\begin{figure}[H]
\centering
\vspace{-.05cm}
    \includegraphics[width=160mm]{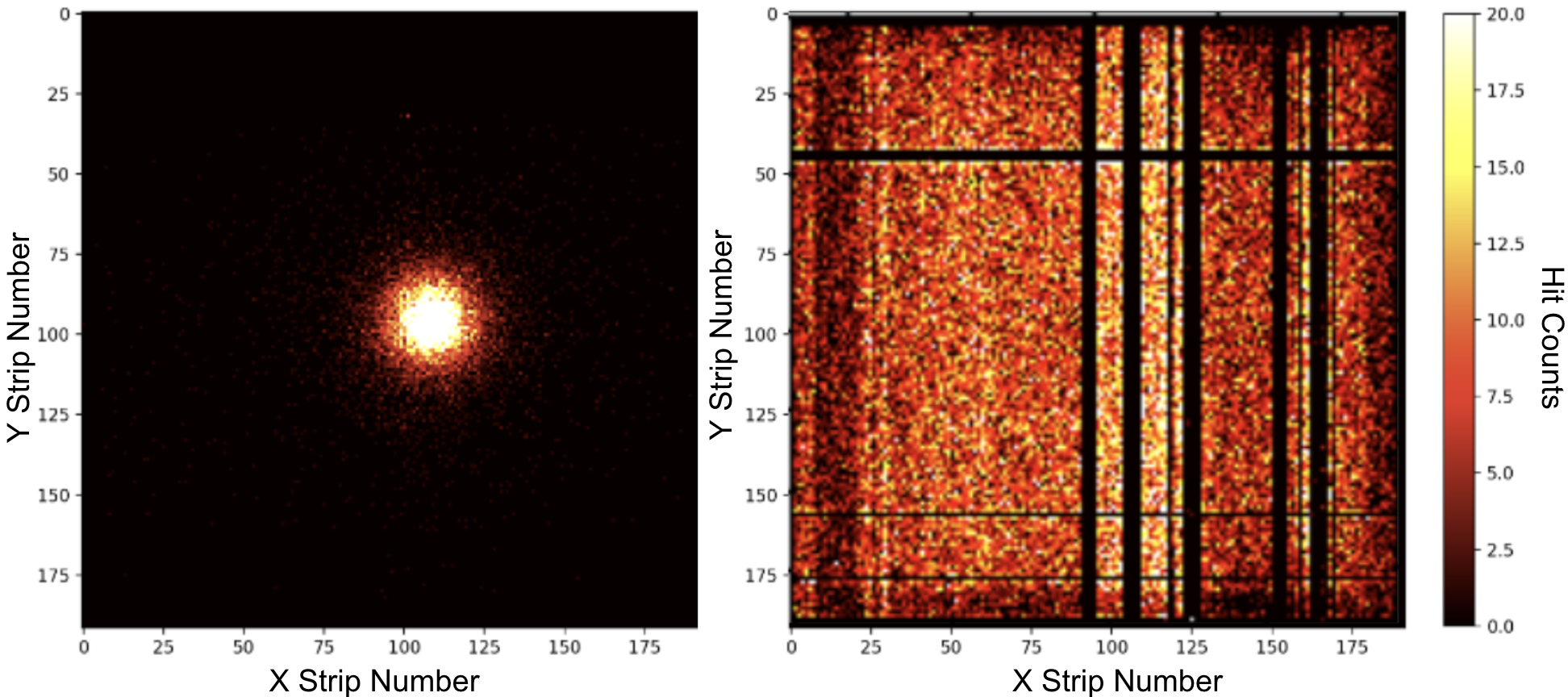}
\caption{(Left) Heat map of Layer 1 generated from a Sr-90 source placed on top of a delrin cover $\sim$0.5 cm above the surface of the DSSD. Short interaction lengths from a beta source like Sr-90 result in a hot spot coincident with the source location. This allows for a quick check of general detector performance and a confirmation that the ASIC channel mapping is correct in the data acquisition software. (Right) Heat map of Layer 1 generated from a Na-22 source placed above the detector. Using a gamma source like Na-22, the entire detector is illuminated to quickly show which ASIC channels are masked or have anomalous count rates.}
\label{fig:Sr90_Heatmap}
\end{figure}

Setting optimized thresholds and mask settings is essential in preparing the detectors for source-testing. In source-testing, radioactive sources are placed above the DSSD, and data is collected for a minimum of one minute. The data from all strips is then accumulated and checked to determine if a clear photo-peak is present (more information of how these photo-peaks are obtained is detailed in § 3.1). A heat map of the detector hits is also created for a visual confirmation of the active area of the detector. For beta sources, the heat map should show a circular gradient centered around where the source was placed for a quick check of general detector performance as well as confirmation of correct ASIC channel mapping (see Fig.~\ref{fig:Sr90_Heatmap}). For gamma-ray sources, the heat map should show all unmasked strips illuminated. If a detector passes these visual tests, and a photo-peak is clearly identified from a gamma-ray source, then the layer is ready for integration into the full detector stack, which entails physically stacking and securing the layers on top of each other. Once the layers are integrated, they then are ready for energy calibrations. 

\subsection{Data Acquisition}

Each of the 10 Tracker layers can operate independently and are controlled through the connection with the FPGA on each DBE. In normal operation with the rest of the ComPair detector subsystems, the Tracker layers are connected to the Trigger Module, which checks for coincidence between layers and other detectors systems to enable the system readout, which drastically suppresses noise in the data. Different trigger conditions are set based on the testing and operating configuration. The ``normal mode" for operation requires either two hits in the Tracker, one on each side of the detector, or coincidence between any two of the Tracker, CZT calorimeter, and CsI calorimeter. If either condition is met, all layers of the Tracker, as well as all other subsystems are triggered to read out data. This mode was used for full-system calibrations, as well as taking beam-test and balloon flight data. When the Tracker was operating without other subsystems or the Trigger Module, each layer was read out independently. In this mode, if a single strip read out above threshold, it would send a trigger that would in turn read out all the strips in that layer. This was the mode used for calibrating the Tracker.

As mentioned above, when one channel in the Tracker measures a signal above threshold, all of the channels in the layer are then readout. A hold time of 2 $\mu$s is used to hold the pulse height information while the firmware logic is confirmed for readout. Occasionally, it was found that an ASIC would freeze during readout, which caused the full layer to lock up waiting for the event readout to fully finish. To avoid this error, a readout timeout of 300 $\mu$s was set in the Tracker firmware so the system could move on and continue taking data.

As data is being read out, the Tracker has stand alone python analysis tools to examine the data. These tools allow for real-time monitoring of raw rates, individual channel ADC values, and housekeeping. After the data is recorded, it is then run through an off-line pipeline to filter and calibrate events before combining with data from the other subsystems. This pipeline includes a strip pairing algorithm that resolves each interaction, detailed below.

\begin{figure}[H]
\centering
    \includegraphics[width=160mm]{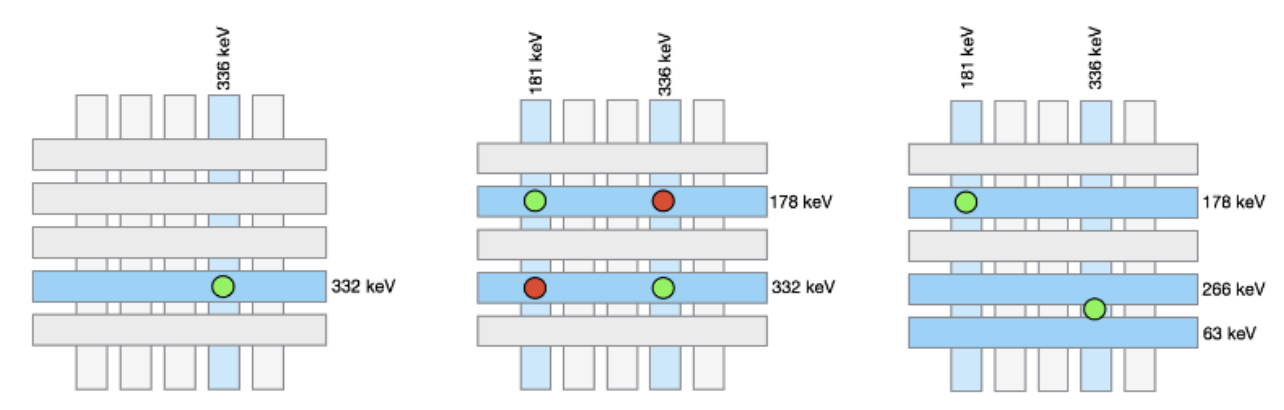}
\caption{(Left) For single interactions, the position is found by taking the intersection of the strips, shown in green. (Center) For multiple interactions, multiple solutions exist, shown in red and green. In order to resolve the positions of the interactions, the strips that are closest in energy are paired together, thus the green circles are used in this case. (Right) When charge is shared between two adjacent strips, the position and energy must be resolved first before resolving multiple interactions. This is done by summing the energies of neighboring interactions and using a weighted average to obtain the position of the hit. Figures from Ref. \citeonline{Sleator_2019}.}
\label{fig:strip_pairing}
\end{figure}

When a charge is induced in a detector strip, it is denoted as a strip hit. In the simplest case of a single interaction, where a particle generates a strip hit on either side of the detector, the position is simply the intersection of the two strips and the p-side strip energy is used as the energy of the interaction (since it is more accurate; see § 3.1). For multiple coincidence interactions in a detector, where there are multiple solutions to the positions of the interactions, n-side and p-side strips are paired together based on the measured energy (Fig.~\ref{fig:strip_pairing}). For instances of charge-sharing, where two neighboring strips share the energy from an interaction, the energy of the neighboring strip hits is summed and the position is taken as a weighted average between the two strips. It is noted that this is done before resolving multiple interactions to more accurately match the n-side and p-side energies.

\section{INDIVIDUAL LAYER PERFORMANCE}

\subsection{Calibrations}

\begin{figure}[H]
  \centering
  \begin{minipage}[b]{0.46\textwidth}
    \includegraphics[width=\textwidth]{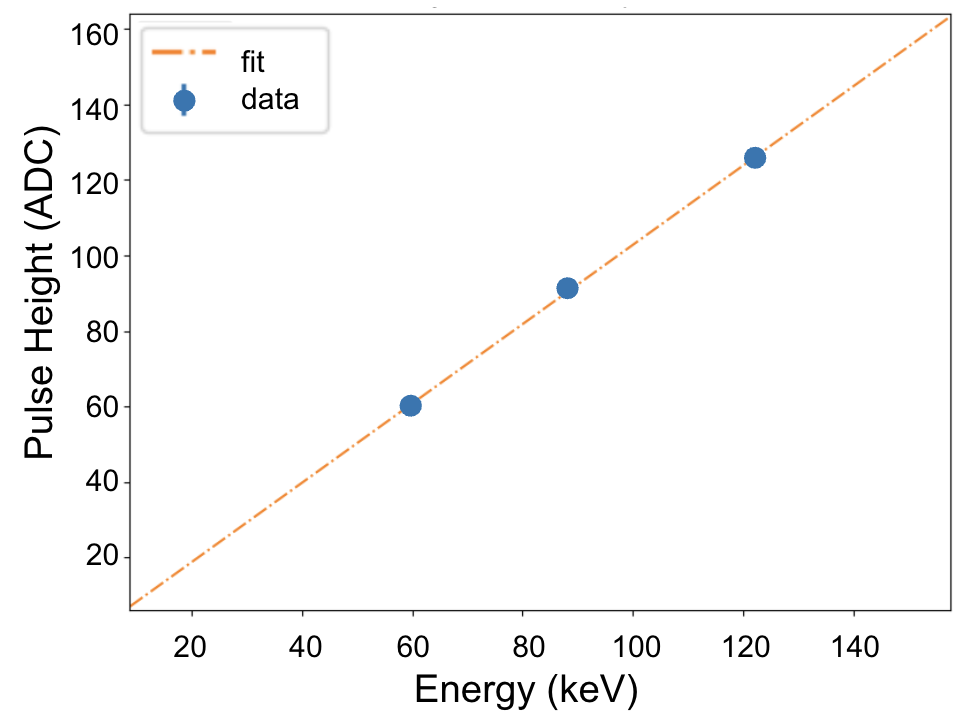}
    \subcaption{Linear fit of 3 radioactive sources}
    \label{fig:EnergyCalib}
  \end{minipage}
  \hfill
  \begin{minipage}[b]{0.51\textwidth}
    \includegraphics[width=\textwidth]{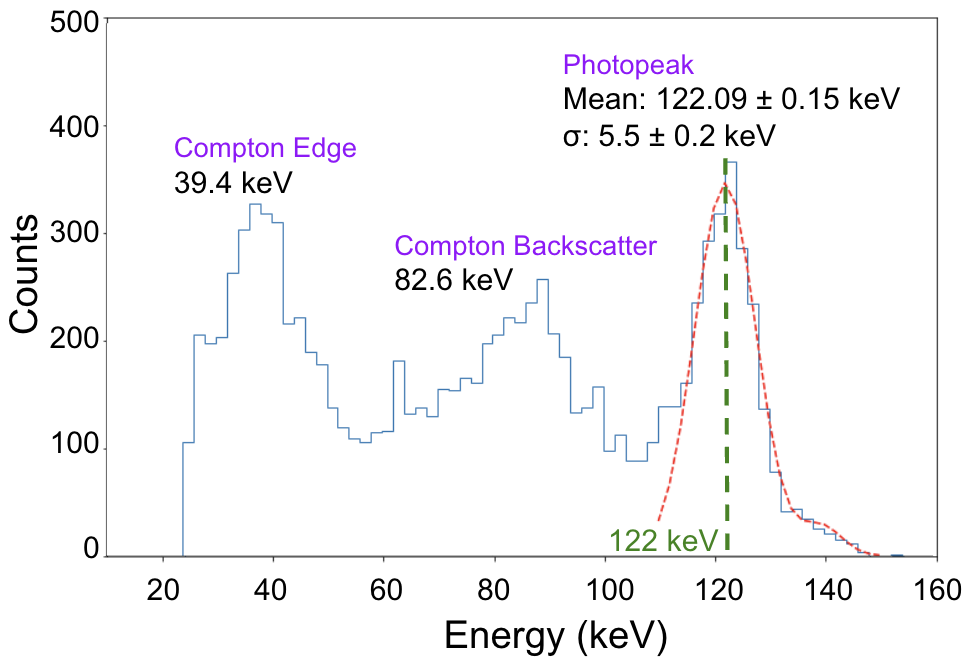}
    \subcaption{Energy Spectrum of Co-57}
    \label{fig:EnergySpectrum}
  \end{minipage}
  \caption{(a) Plot of ADC values versus known emission line energies of three radioactive sources from Strip 89 on the p-side of Layer 1. A linear fit is used to find the relation between ADC and energy. The error bars from the centroid fit value are too small to be visible on the plot. (b) The Co-57 spectrum of Strip 89 on the p-side of Layer 1 after energy calibration shows the 122 keV photopeak and a faint 136 keV photopeak. This is fit with a double Gaussian line resulting in a single strip energy resolution of 5.5 $\pm$ 0.2 keV $\sigma$ at 122 keV. The known decay emission line of 122 keV is plotted in green.}
  \label{fig:FullEnergySpectrum}
\end{figure}

To perform energy calibrations for the DSSD Tracker, radioactive sources with known decay emission lines are used. The sources span an energy range of 14 keV to 136 keV (Table~\ref{tab:Sources}), which fall on the lower side of the Tracker's dynamic range of $\sim$20 keV to $\sim$700 keV. This is due to the fact that for thin silicon detectors at higher energies, Compton scattering is the dominant photon interaction process, and would be difficult to use for calibrations since no photopeak would be present. However, for the energies that the selected radioactive sources are at, the photoelectric effect is prominant. These sources are placed directly above the detector, and radiated photons interact with the silicon in the DSSD via Compton scattering and the photoelectric effect. The result from the photoelectric effect is photoelectrons, which then ionize in the silicon, creating electron-hole pairs. After a couple hours run with a radioactive source, each strip typically collects enough hits to form a preliminary spectrum. This spectrum is a histogram of ADC values for all the detected counts in the run. In the ideal case, the spectrum would have a narrow line at a certain ADC value corresponding to the energy of the emission line of the source. However, due to the high leakage current and capacitance in the detector, the photopeak is broadened in the actual spectrum. A Gaussian distribution is fit to the photopeak and the centroid ACD value is mapped to the energy of the source and the FWHM of the fit corresponds to the resolution of the strip. Since the response of each strip and ASIC channel can vary, it is necessary to measure and calibrate each one individually. This process is then repeated with several sources emitting at various energies. 3 radioactive sources in total are used per strip, and a simple linear relation is used to describe the relation between ADC values and energy (Fig.~\ref{fig:EnergyCalib}). This relation can then be used to map values from the ADC space to the energy space and to generate a spectrum of a source in keV (Fig.~\ref{fig:EnergySpectrum}). Work is currently underway to incorporate higher energy data points into these calibrations to make the Tracker's high energy response more accurate. For example, while the 662 keV photopeak of Cs-137 cannot be resolved by the Tracker, the Compton edge at $\sim$490 keV could be fit and used.

 
\begin{table}[ht]
\caption{Radioactive sources used to calibrate the DSSD Tracker on {\it ComPair} \cite{radioactive_sources}.}
\label{tab:Sources}
\begin{center} 
\begin{tabular}{|l|l|lll}
\hline
Nuclides & Half-life & \multicolumn{1}{l|}{Line Origin} & \multicolumn{1}{l|}{Line Energies (keV)} & \multicolumn{1}{l|}{Transition Probability} \\ \hline
\multirow{3}{*}{Co-57}  & \multirow{3}{*}{272 days}    & \multicolumn{1}{l|}{$\gamma$}        & \multicolumn{1}{l|}{14.41}               & \multicolumn{1}{l|}{0.09}                 \\ \cline{3-5} 
                        &                              & \multicolumn{1}{l|}{$\gamma$}        & \multicolumn{1}{l|}{122.06}              & \multicolumn{1}{l|}{0.86}                  \\ \cline{3-5} 
                        &                              & \multicolumn{1}{l|}{$\gamma$}        & \multicolumn{1}{l|}{136.47}              & \multicolumn{1}{l|}{0.11}                 \\ \hline
\multirow{3}{*}{Cd-109} & \multirow{3}{*}{464.4 days}  & \multicolumn{1}{l|}{Ag-SumK$\alpha$}     & \multicolumn{1}{l|}{22.10}                & \multicolumn{1}{l|}{0.84}                  \\ \cline{3-5} 
                        &                              & \multicolumn{1}{l|}{AgSumK$\beta$}      & \multicolumn{1}{l|}{24.90}                  & \multicolumn{1}{l|}{0.18}                 \\ \cline{3-5} 
                        &                              & \multicolumn{1}{l|}{$\gamma$}        & \multicolumn{1}{l|}{88.03}               & \multicolumn{1}{l|}{0.04}                \\ \hline
Am-241                  & 432.2 years                  & \multicolumn{1}{l|}{$\gamma$}        & \multicolumn{1}{l|}{59.40}                & \multicolumn{1}{l|}{0.36}                  \\ \hline

\end{tabular}%
\end{center}
\end{table}


It is noted that for these calibration runs, the Tracker was not connected to the Trigger Module since event coincidence between Tracker layers was not relevant. Instead, for any hit above threshold, all strips in the detector were read out. This data was then fed into the calibration pipeline. The Tracker calibration pipeline was automated using custom python code that automatically fit the photo peaks for each strip. However, the variance between strips would sometimes lead to improper fits so all outliers were manually checked and often resolved with manual fits. 

\begin{figure}[H]
\centering
    \includegraphics[width=100mm]{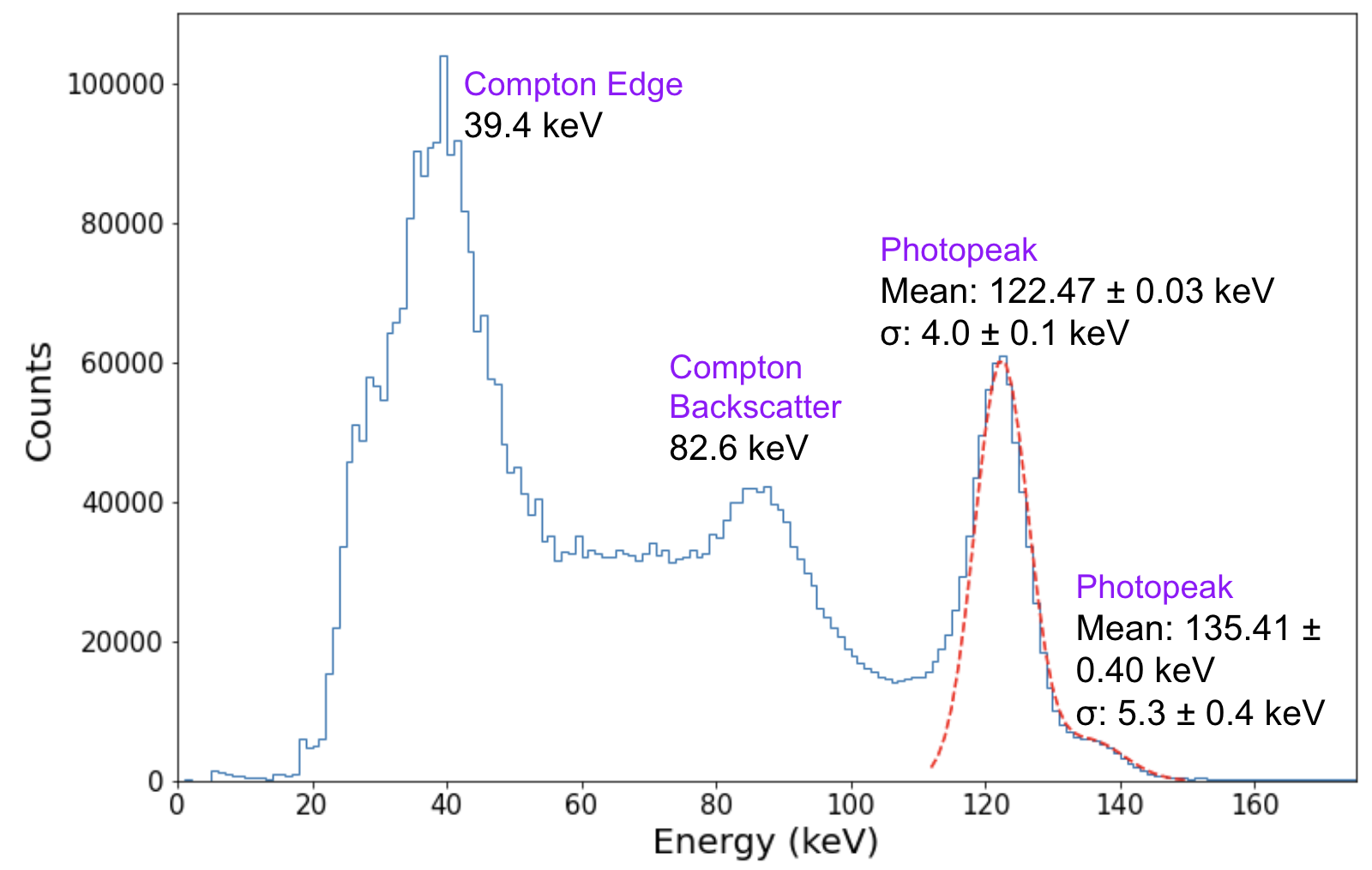}
\caption{A Co-57 spectrum generated by summing the energy-calibrated spectra of each strip on Layer 1. The 122 and 136 keV photopeaks were simultaneously fit with Gaussian lines resulting in energy resolutions of 4.0 $\pm$ 0.1 keV $\sigma$ and 5.3 $\pm$ 0.4 keV $\sigma$, respectively. The Compton edge was calculated to be around 39 keV via Eq. (\ref{eq:Compton}), with back scatter from detectors in other layers at around 82 keV.}
\label{fig:Co57}
\end{figure}

\begin{figure}[H]
  \centering
  \begin{minipage}{0.51\textwidth}
    \begin{tabular}{|c|c|c|} 
        \hline
        Layer & n-side FWHM (keV) & p-side FWHM (keV) \\ 
        \hline
        0 & 36.56 $\pm$ 8.05 & 12.88 $\pm$ 0.83\\
        1 & 15.77 $\pm$ 1.06 & 10.25 $\pm$ 0.28\\ 
        2 & 17.95 $\pm$ 0.34 & 9.76 $\pm$ 0.21\\
        3 & 37.92 $\pm$ 2.07 & 10.36 $\pm$ 0.32\\ 
        4 & 24.47 $\pm$ 0.52 & 11.03 $\pm$ 0.46\\
        5 & 19.37 $\pm$ 0.49 & 13.05 $\pm$ 0.40\\ 
        6 & 15.26 $\pm$ 0.56 & 10.38 $\pm$ 0.28\\
        7 & 15.64 $\pm$ 0.82 & 11.75 $\pm$ 0.62\\
        8 & 21.58 $\pm$ 1.53 & 14.70 $\pm$ 1.38\\
        9 & 18.18 $\pm$ 0.49 & 10.28 $\pm$ 0.29\\
        \hline
    \end{tabular}
    \vspace{1.2cm}
    \subcaption{Co-57 122 keV Photopeak Widths}
    \label{tab:photopeak_widths}
  \end{minipage}
  \begin{minipage}{0.48\textwidth}
    \includegraphics[width=\textwidth]{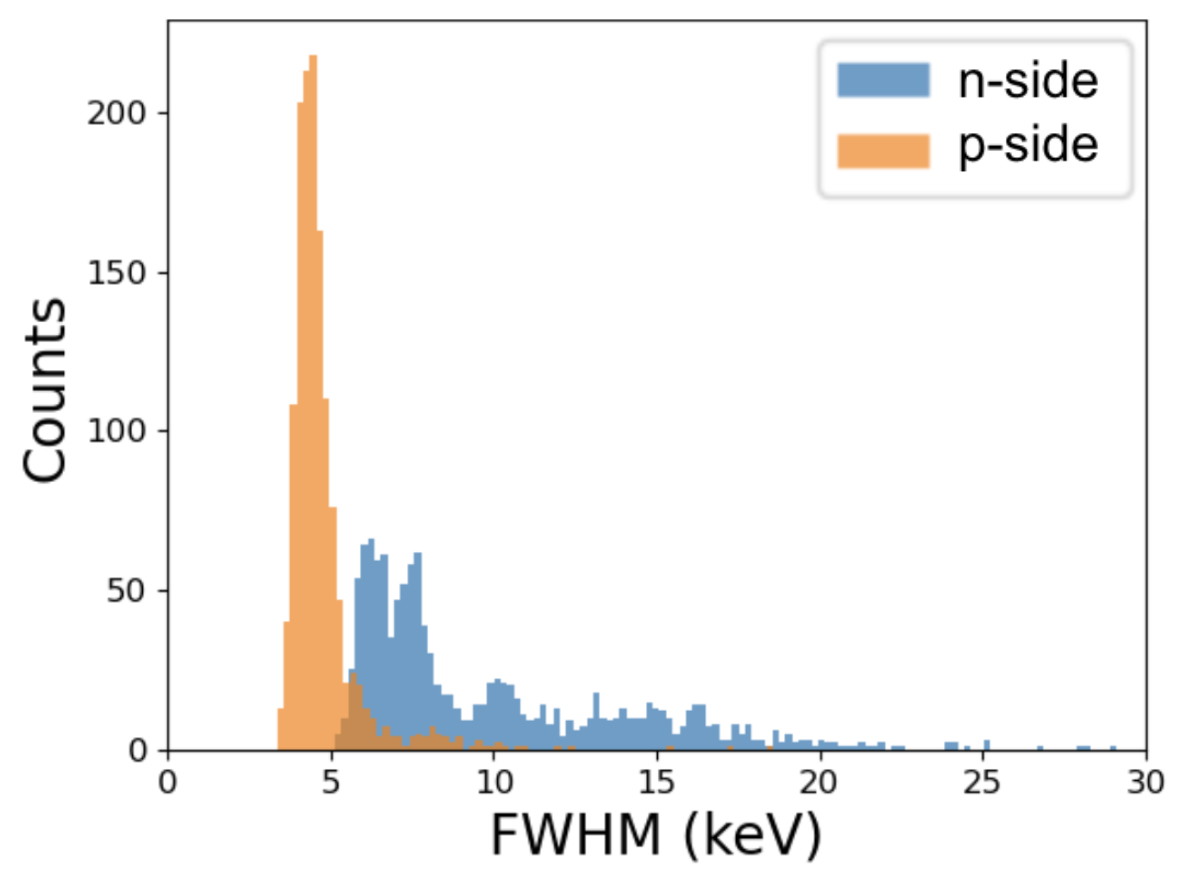}
    \vspace{-0.5cm}
    \subcaption{Tracker strip resolution by side}
    \label{fig:StripEnergyRes}
  \end{minipage}
  \caption{(a) For each side of each Tracker layer, the Co-57 122 keV photopeak was fit with a Gaussian line. The full-width half-max (FWHM) of the Gaussian was then recorded representing the energy resolution per side per layer. (b) For each Tracker strip, the Co-57 122 keV photopeak was fit with a Gaussian line. The p-side (orange) energy resolution of each detector was found to be globally better.}
\end{figure}

A full-layer energy spectrum is generated by applying the calibration file to the source run (Fig.~\ref{fig:Co57}).  A Gaussian was then fit to the photopeak to confirm a consistent response among all the strips as well as to understand the energy resolution of the layer (Fig.~\ref{tab:photopeak_widths}). For a more in-depth analysis of the detectors, the same process was repeated on the energy spectra generated from individual strips (Fig.~\ref{fig:StripEnergyRes}).

\begin{figure}[H]
  \centering
  \begin{minipage}[b]{0.43\textwidth}
    \includegraphics[width=\textwidth]{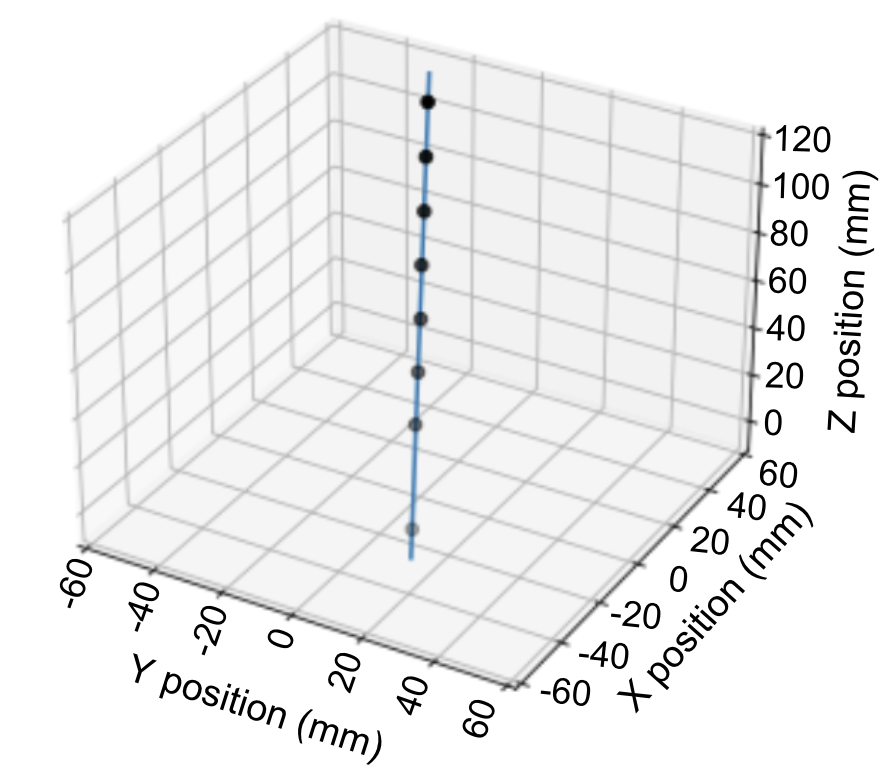}
    \subcaption{Muon Track}
    \label{fig:MuonTrack}
  \end{minipage}
  \hfill
  \begin{minipage}[b]{0.56\textwidth}
    \includegraphics[width=\textwidth]{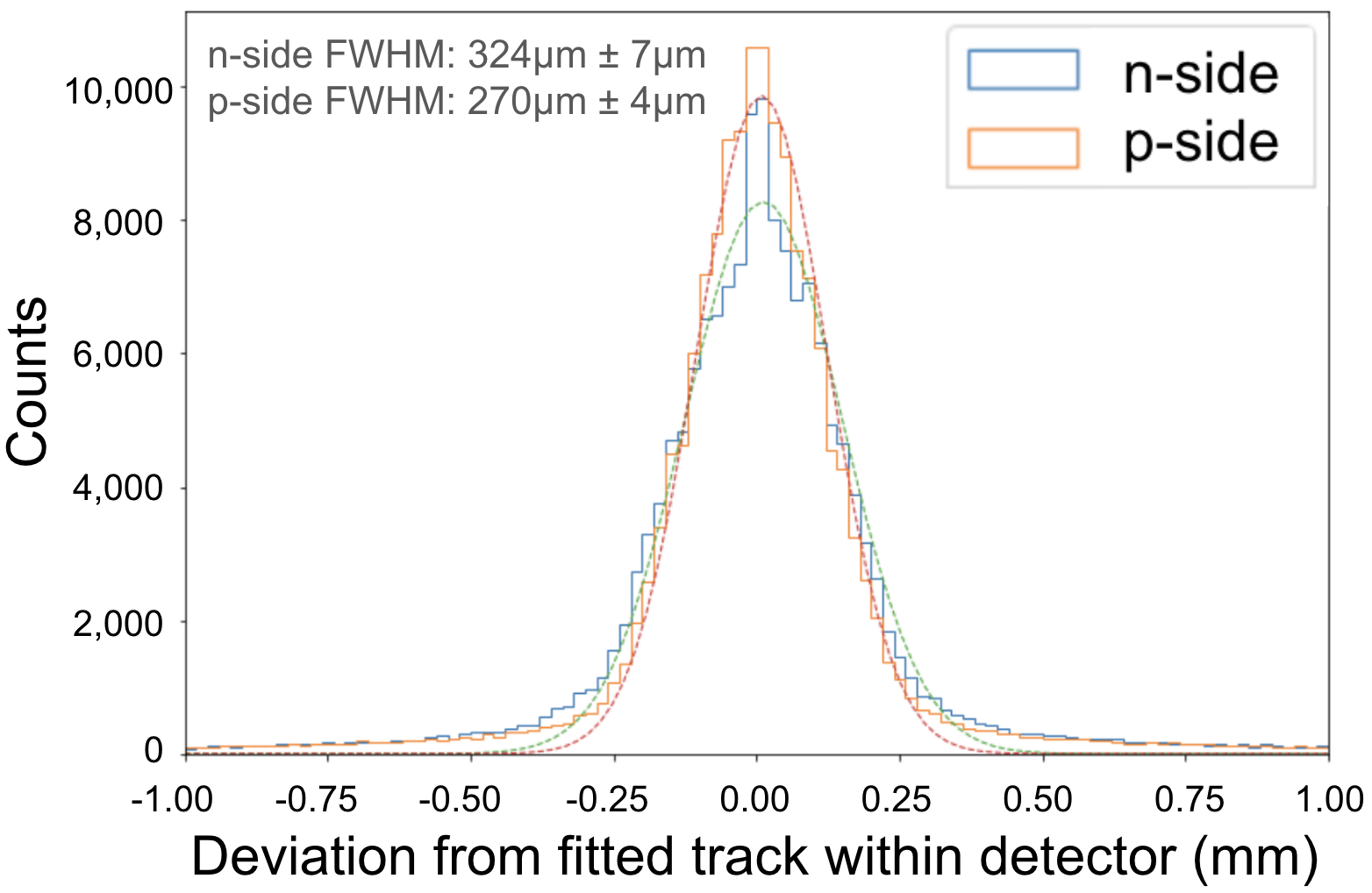}
    \subcaption{Tracker position resolution}
    \label{fig:TKR_PosRes}
  \end{minipage}
  \caption{(a) Muon track recorded in 8 layers of the Tracker, fit with a line. (b) Distribution of deviations measured from recorded hits to fitted muon tracks.}
\end{figure}

The position resolution and alignment of the Tracker was quantified using muon tracks. With the Tracker layers all stacked on top of each other, data was recorded for any event that triggered both sides of at least 1 detector. For events that contained hits in at least 5 layers, a linear fit was performed on the muon tracks. Due to occasional coincident hits in a detector, some hit positions were improperly recorded. To account for this, any hit that had over a 5cm deviation from the fitted track was removed, and the remaining layers were refit with a straight line (Fig.~\ref{fig:MuonTrack}). The deviation between the recorded hit position and the fitted track's intersection in a given layer was recorded. The distribution of deviations was then plotted and fit with a Gaussian (Fig.~\ref{fig:TKR_PosRes}), where the FWHM of was found to be 324 $\mu$m $\pm$ 7 $\mu$m and 270 $\mu$m $\pm$  4$\mu$m for the n-sides and p-sides, respectively. This exceeded the expected FWHM of 346.7 $\mu$m which was calculated based on Molière scattering \cite{Moliere}. This is attributed to the multihit and charge sharing corrections made in the Tracker pipeline (§ 2.4).

\section{FULL STACK PERFORMANCE}

\subsection{Beam Test}

\begin{figure}[H]
  \centering
  \begin{minipage}[b]{0.45\textwidth}
    \includegraphics[width=\textwidth]{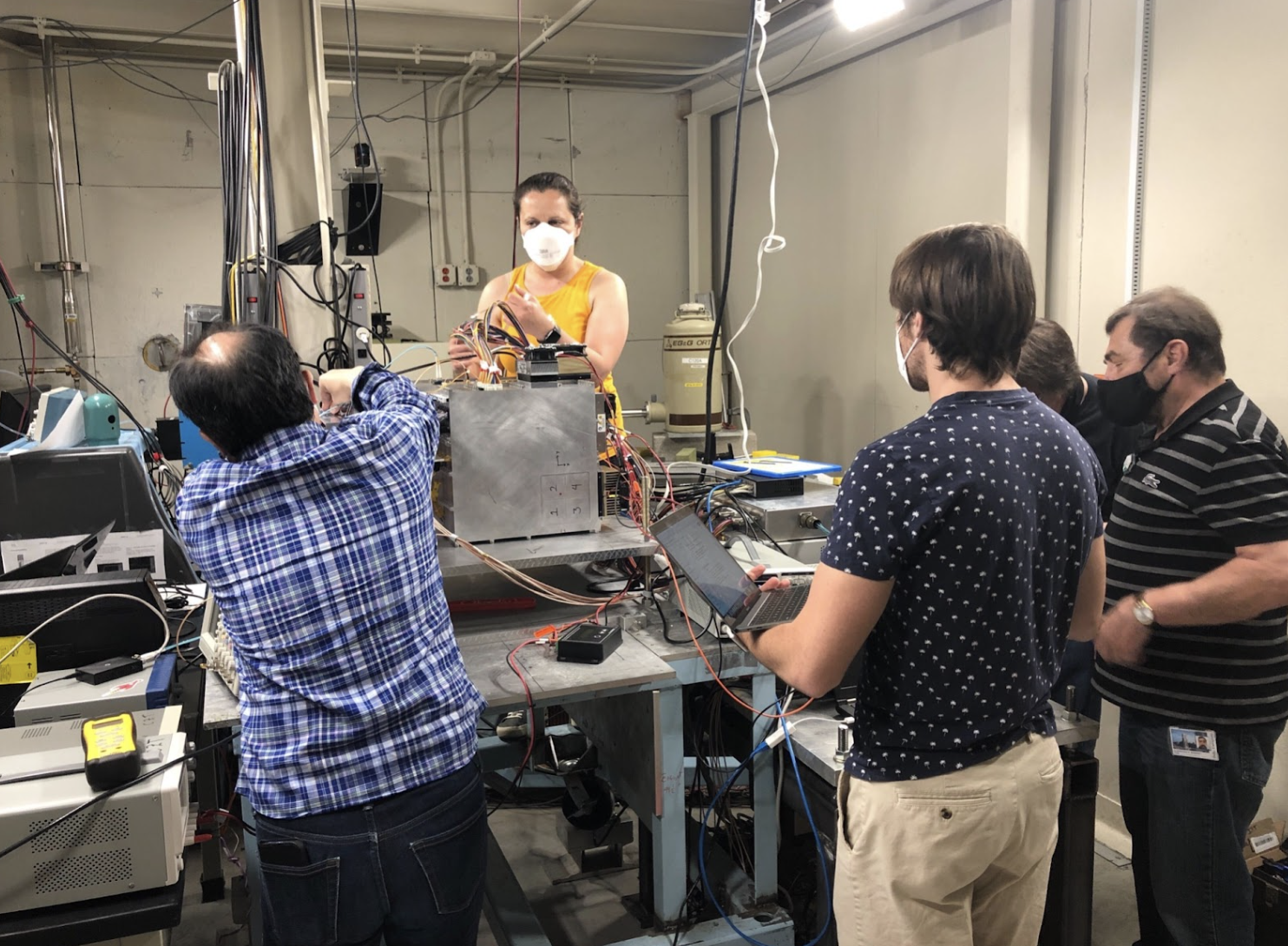}
    \vspace{0.22cm}
    \subcaption{{\it ComPair} at the HIGS facility.}
    \label{fig:HIGS_img}
  \end{minipage}
  \hfill
  \begin{minipage}[b]{0.54\textwidth}
    \includegraphics[width=\textwidth]{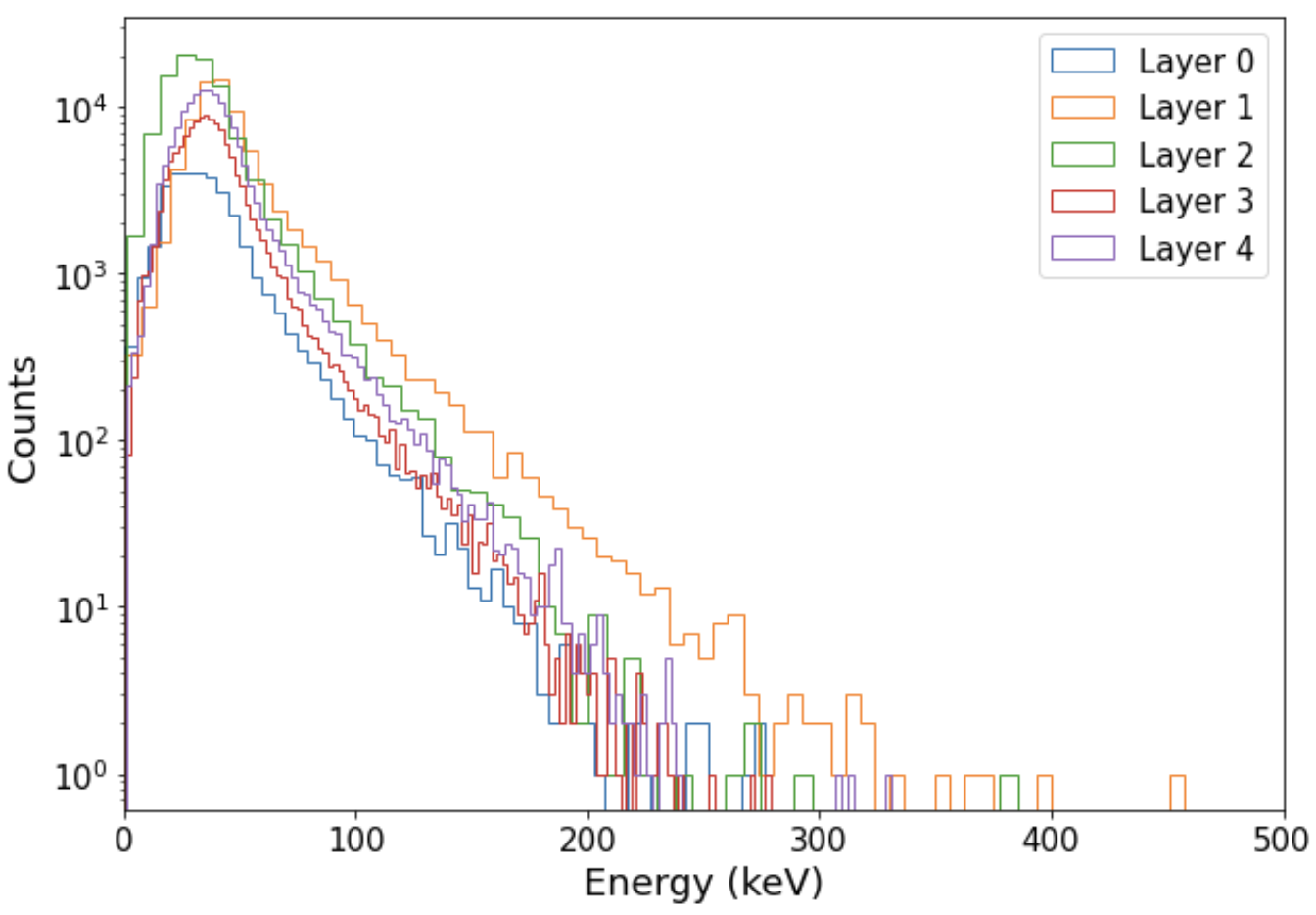}
    \subcaption{Individual Tracker layer spectra of the 15 MeV beam.}
    \label{fig:Beam_Spectra}
  \end{minipage}
  \caption{(a) The {\it ComPair} team sets up the instrument at the HIGS facility. Since the beam came in parallel to the ground, {\it ComPair} was placed on its side so the beam was perpendicular to the detectors. (b) Spectra for each layer of the Tracker from a 15 MeV gamma-ray beam run. At this energy, many photons will undergo pair-production, and the electron-positron pairs will experience a constant energy loss as they traverse the silicon detectors.}
\end{figure}

In April of 2022, {\it ComPair} was tested at Duke University's Free Electron Laser Laboratory for testing with their High Intensity Gamma-ray Source (HIGS) Beam (Fig.~\ref{fig:HIGS_img}) \cite{HIGS}. By this time, 5 layers of the Tracker had been tested and integrated with the rest of {\it ComPair}. The purpose of this test was to measure the instrument's performance at higher energies (2 MeV to 25 MeV), unachievable with radioactive sources. These energies surpass the 1.02 MeV threshold for pair-conversion to take place, therefore the beam test would help validate {\it ComPair}'s performance in the regime where pair-conversion is prominant.

Over the course of a week, a gamma-ray beam was shot at {\it ComPair} at  energies of: 2, 5.1, 7, 15, and 25 MeV. This was especially crucial for the DSSD Tracker, since one of its main goals is to track electron-positron pairs before they get absorbed in the calorimeters. The electron-positron pairs experience constant energy loss as they traverse each detector in the Tracker, resulting in a continuum spectrum per layer. This can be seen in Fig.~\ref{fig:Beam_Spectra}, which shows spectra from 5 layers of the Tracker generated from a 1 hour 15 MeV gamma-ray run. It is noted that further analysis of the beam test data is underway.

\subsection{TVAC Testing}

In August of 2022, thermal vacuum (TVAC) testing of the Tracker subsystem was performed. The purpose of these tests was to refine thermal models for balloon flight, as well as to test the performance of the Tracker in flight-like conditions. Initially, one layer was tested in the TVAC chamber to understand the operational temperatures of the detectors. The TVAC chamber was pumped down to a pressure of $\sim$1 Torr, and detector performance was tested between $-30~^{\circ}$C and $40~^{\circ}$C using external calibration pulses. It was concluded that the detector was functional from $-20~^{\circ}$C to $35~^{\circ}$C, setting the lower and upper limits for the Tracker's operational temperature. 

\begin{figure}[H]
  \centering
  \begin{minipage}[b]{0.35\textwidth}
    \includegraphics[width=\textwidth]{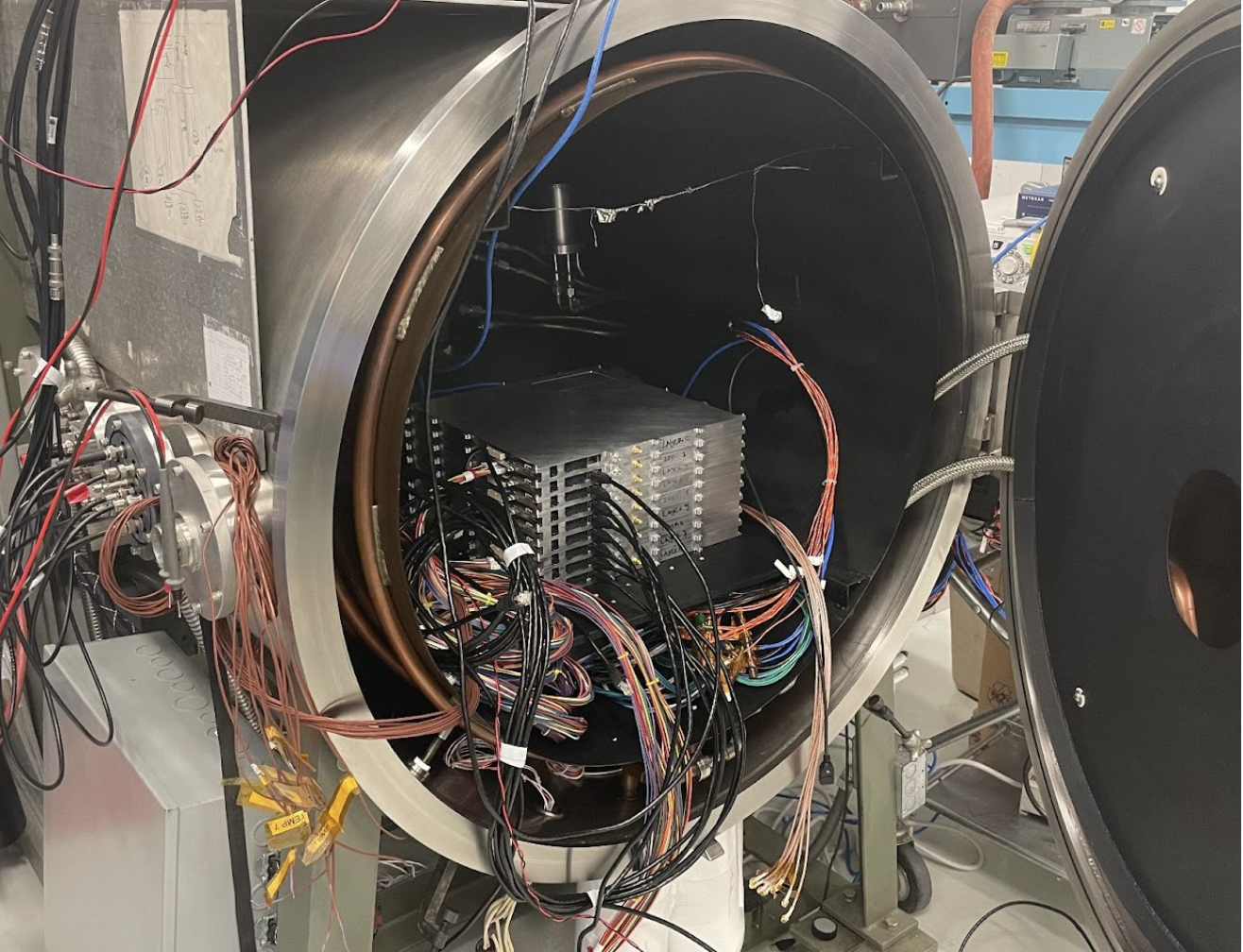}
    \subcaption{Tracker stack in the TVAC chamber}
    \label{fig:Tracker_Vac}
  \end{minipage}
  \hfill
  \begin{minipage}[b]{0.64\textwidth}
    \includegraphics[width=\textwidth]{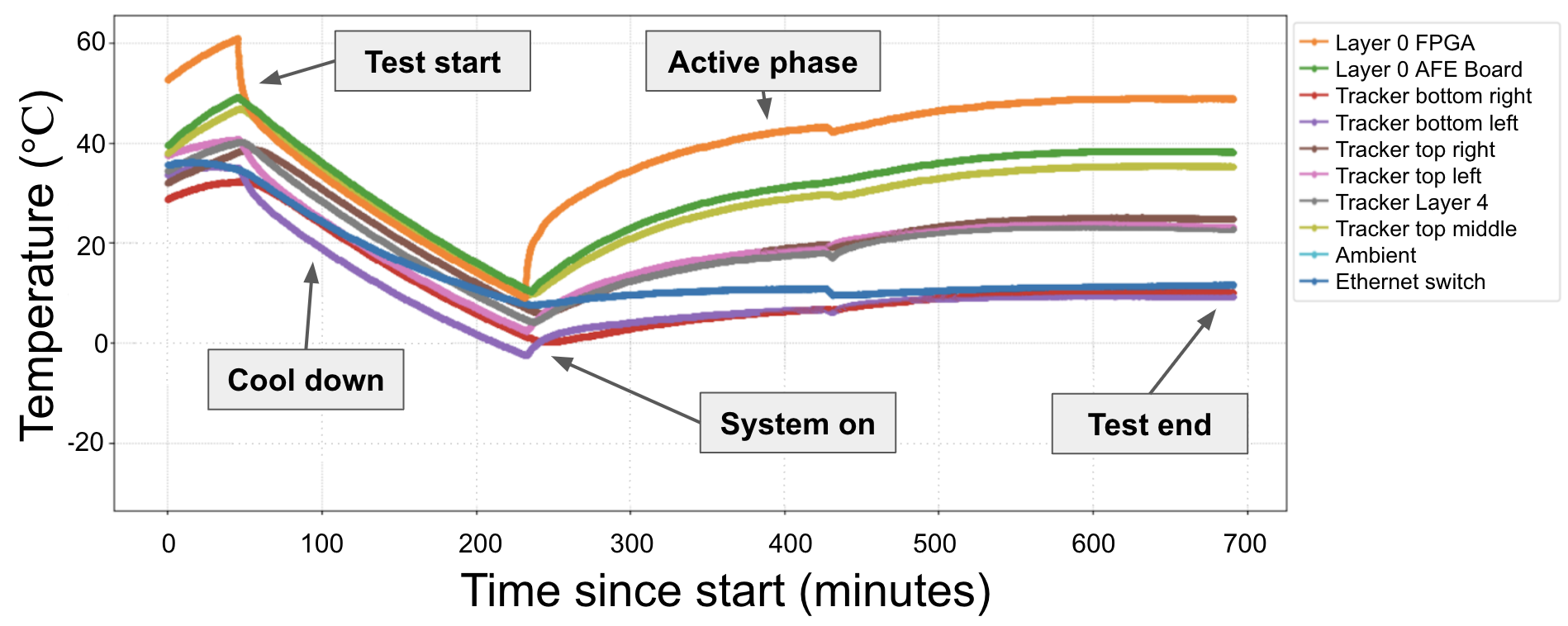}
    \subcaption{Temperatures for 9 layers of the Tracker measured during TVAC}
    \label{fig:TKR_TVAC_Plot}
  \end{minipage}
  \caption{(a) 9 layers of the Tracker stack were tested to benchmark thermal modeling and to understand how fast the Tracker heats up in flight-like conditions. (b) The TVAC test started with the chamber cooling down and the Tracker powered off. The system was turned on after $\sim$4 hours, and at $\sim$7 hours into the test the temperature of the chamber was reduced by $\sim$5$~^{\circ}$C. Soon after, the top of the Tracker surpassed the operational temperature limit. This problem was addressed by adding 4 copper thermal straps to dissipate the heat.}
\end{figure}

9-layers of the Tracker were TVAC tested in December 2022 to better understand the expected operating temperature during flight with all 10 layers powered (Fig.~\ref{fig:Tracker_Vac}). It is noted that the 10th layer was omitted for these tests due to delays in integration. The TVAC test of the stack revealed that the heat produced by the Tracker would pose a problem for the balloon flight, thus necessitating additional ways to dissipate the heat (Fig.~\ref{fig:TKR_TVAC_Plot}). Four copper thermal straps that coupled to the instrument baseplate were fabricated to mitigate this problem before undergoing a second round of TVAC tests.

\begin{figure}[H]
  \centering
  \begin{minipage}[b]{0.44\textwidth}
    \includegraphics[width=\textwidth]{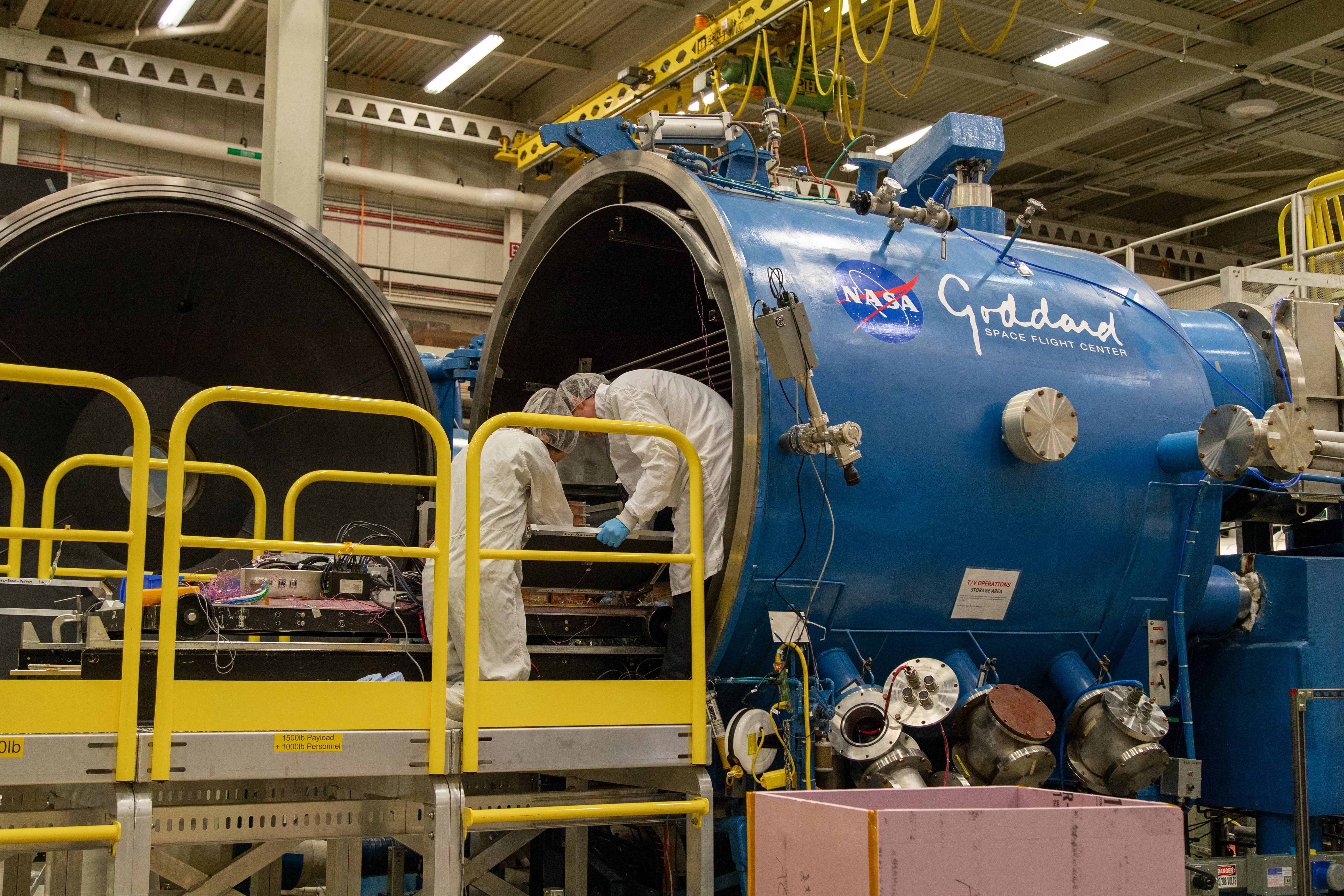}
    \subcaption{{\it ComPair} integration in TVAC chamber}
    \label{fig:TVAC_IMG}
  \end{minipage}
  \hfill
  \begin{minipage}[b]{0.55\textwidth}
    \includegraphics[width=\textwidth]{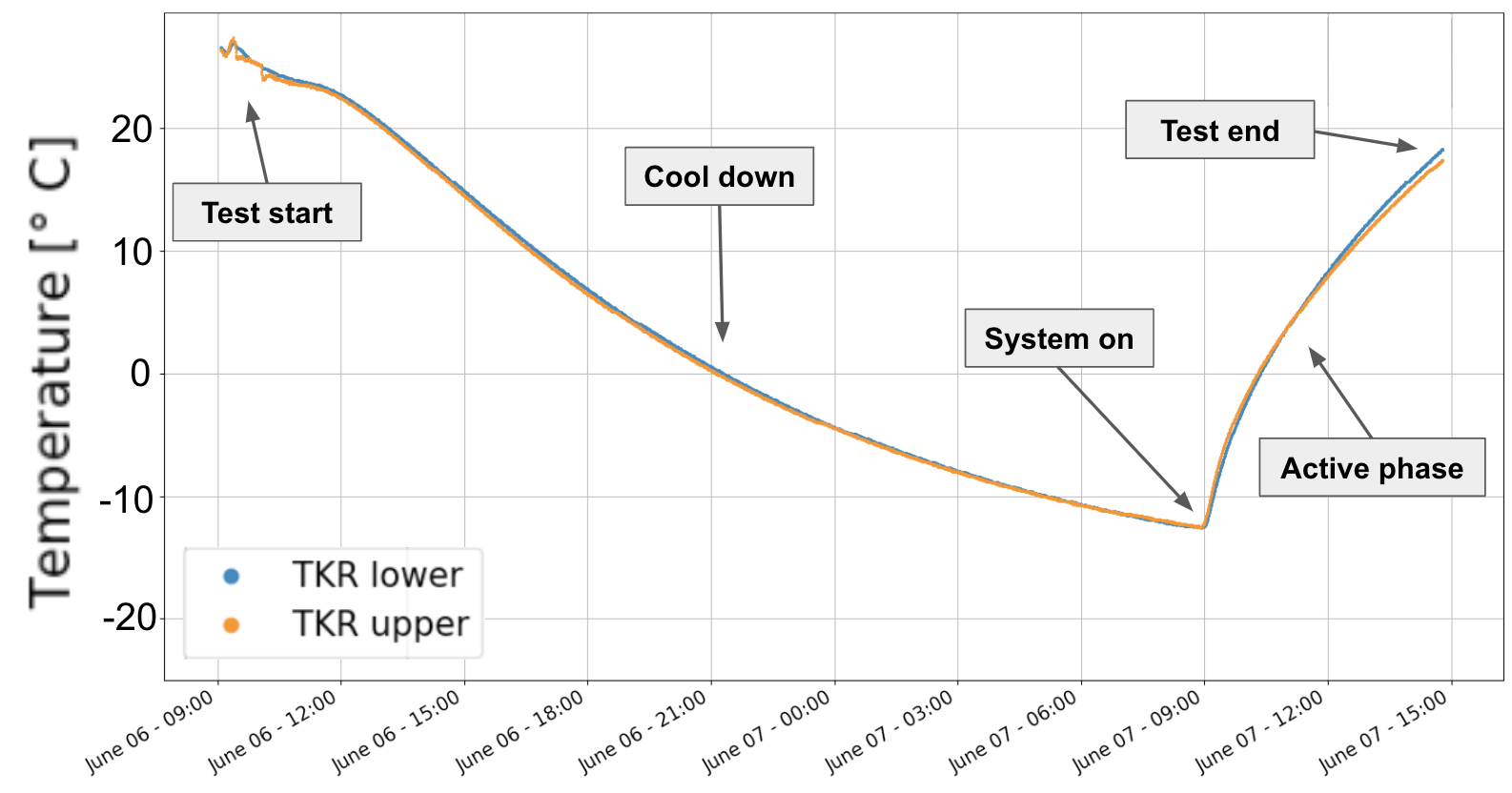}
    \subcaption{Tracker temperatures measured during TVAC}
    \label{fig:TVAC_Plot}
  \end{minipage}
  \caption{(a) Team members on the {\it ComPair} team work on the instrument before it undergoes thermal testing in the TVAC chamber. Credit: NASA/Scott Wiessinger (b) Temperatures measured during TVAC testing show the 24 hour cool-down before the instrument was turned on and tested for an additional 7 hours. Shown in orange and blue are the temperatures of the thermometers placed on the top and bottom of the Tracker enclosure, respectively. }
\end{figure}

In August of 2023, the entire {\it ComPair} instrument and peripherals underwent TVAC testing at NASA GSFC's Integration and Testing facility (Fig.~\ref{fig:TVAC_IMG}) \cite{Valverde_2023}. From the Tracker's perspective, the purpose of these tests were to ensure that the copper thermal straps prevented the subsystem from surpassing the upper operational temperature limit. Initially, the chamber was pumped down to 5 Torr and cooled for 24 hours down to $-20~^{\circ}$C. The instrument and peripherals were then powered on for a 7 hour test in which temperatures were measured and compared to a thermal desktop model. This model was then adjusted to better match the TVAC data. As seen in Fig.~\ref{fig:TVAC_Plot}, the Tracker temperatures stayed within the operational temperature limit, thus confirming that the thermal straps were sufficient for flight.

\subsection{Balloon Flight}

On August 27, 2023, {\it ComPair} took flight out of Fort Sumner, New Mexico, drifting up to a height of $\sim$132,000 ft or $>$99.9\% above the Earth's atmosphere \cite{Pekar_2023}. As a piggyback to the {\it GRAPE} mission \cite{Karla_2022}, it collected data during the $\sim$2.5 hour ascent and during its $\sim$3 hours of float at altitude. The purpose of {\it ComPair's} balloon flight was to observe the gamma-ray background and to test the instruments performance in a flight-like environment. Subsystem event rates, detector temperatures, Tracker bias voltages, and CZT and Tracker high voltage currents were all telemetered down during flight for real-time diagnostics. More information on the balloon flight can also be found in the companion paper titled {\it The ComPair Balloon Instrument and Flight}\cite{Smith_2024}.

Leading up to launch, problems with Tracker Layer 8 arose. Initially, during pre-flight calibrations in Fort Sumner, the layer's rates were significantly higher than the other layers. Further analysis revealed that the noise floor on Layer 8 was much higher than expected, and even with high thresholds the trigger rate was not controllable. After several attempts to fix it, Layer 8 was ultimately disabled and did not record data during flight in order to prevent it from corrupting the rest of the data.

\begin{figure}[H]
  \centering
  \begin{minipage}[b]{0.54\textwidth}
    \includegraphics[width=\textwidth]{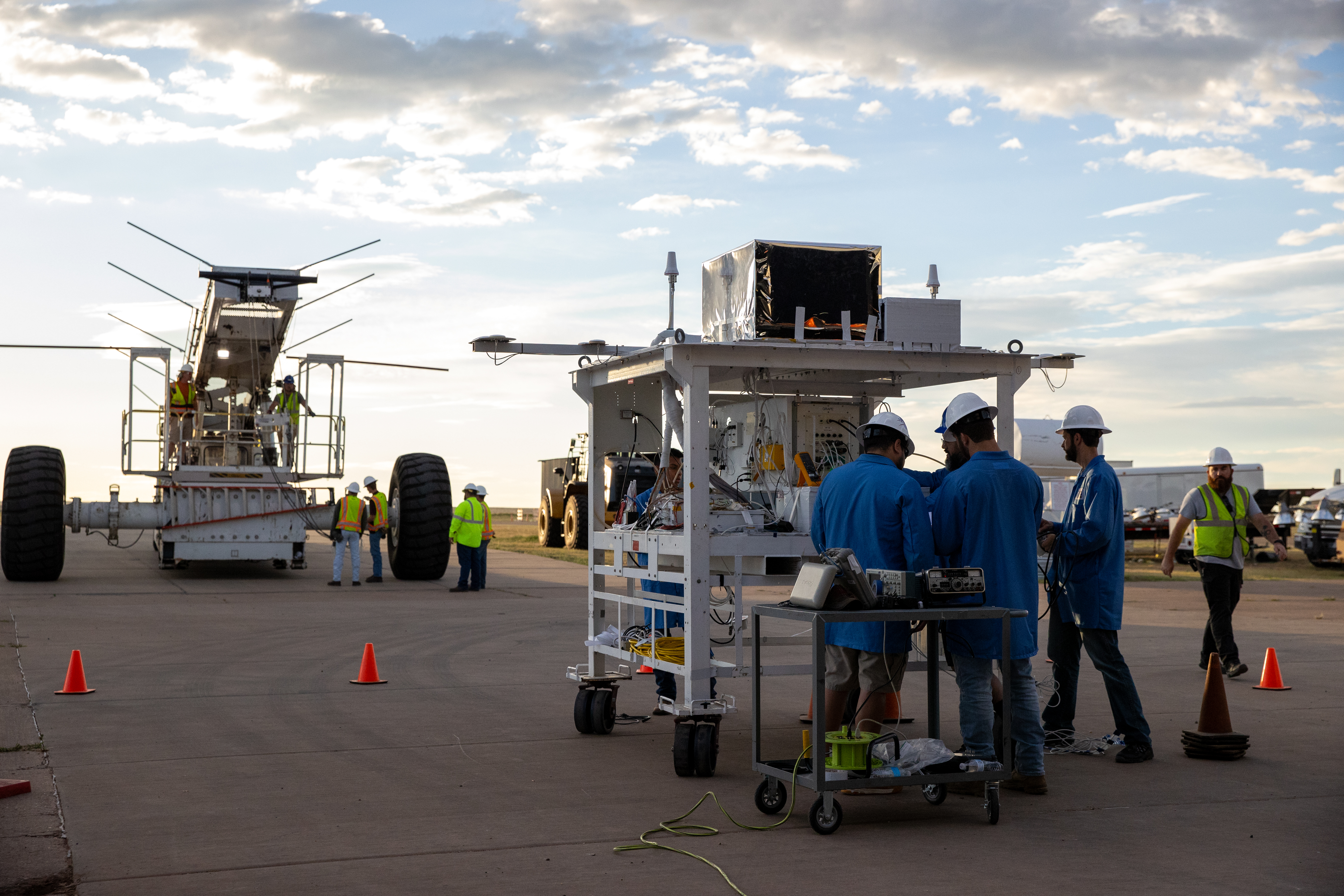}
    \vspace{0.15cm}
    \subcaption{The {\it GRAPE-ComPair} gondola}
    \label{fig:Balloon}
  \end{minipage}
  \hfill
  \begin{minipage}[b]{0.45\textwidth}
    \includegraphics[width=\textwidth]{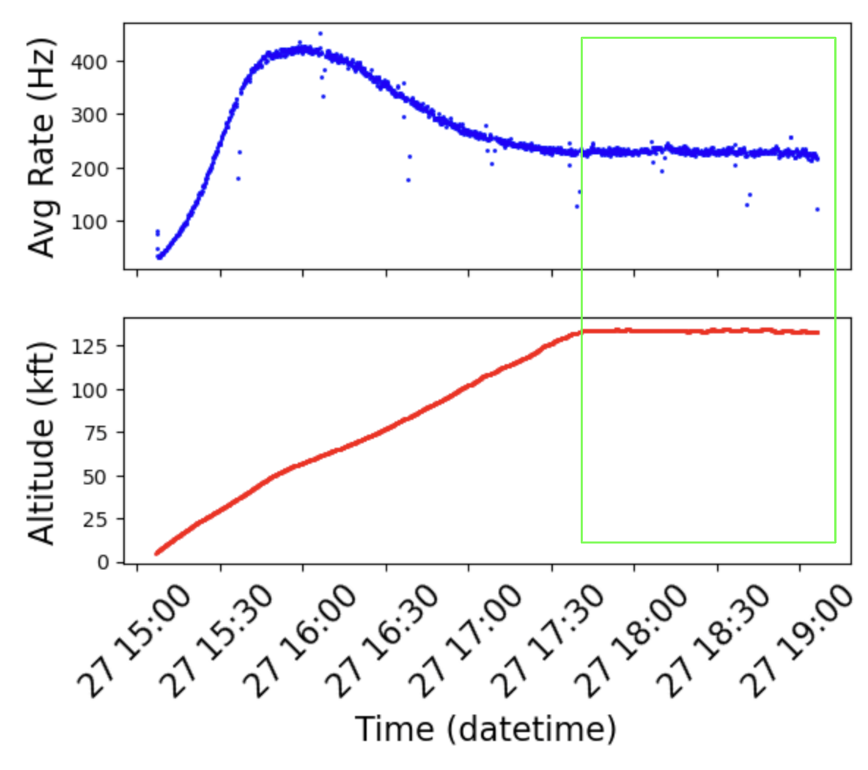}
    \subcaption{Tracker event rate}
    \label{fig:TKR_Flight}
  \end{minipage}
  \caption{(a) Members of NASA's Columbia Scientific Balloon Facility (CSBF)  prepare the {\it GRAPE-ComPair} gondola for a launch attempt. In the background on the left is the launch vehicle that carried the gondola to the launch location. (b) The Tracker event rate as a function of time, plotted above the altitude as a function of time. There is a peak in count rate at the Regener-Pfotzer maximum \cite{Chughtai_2022}, before plateauing once reaching float altitude (boxed in green).}
\end{figure}

 Figure \ref{fig:Balloon} shows the {\it GRAPE-ComPair} gondola being staged for a launch attempt. During the balloon's ascent, the Tracker event rate peaked at $\sim$70,000 ft, corresponding to the Regener-Pfotzer maximum (i.e. where the highest concentration of particle decay occurs) (Fig.~\ref{fig:TKR_Flight}) \cite{carlson_2014}. The event rate then plateaued once reaching float.

\begin{figure}[H]
  \centering
  \begin{minipage}[b]{0.495\textwidth}
    \includegraphics[width=\textwidth]{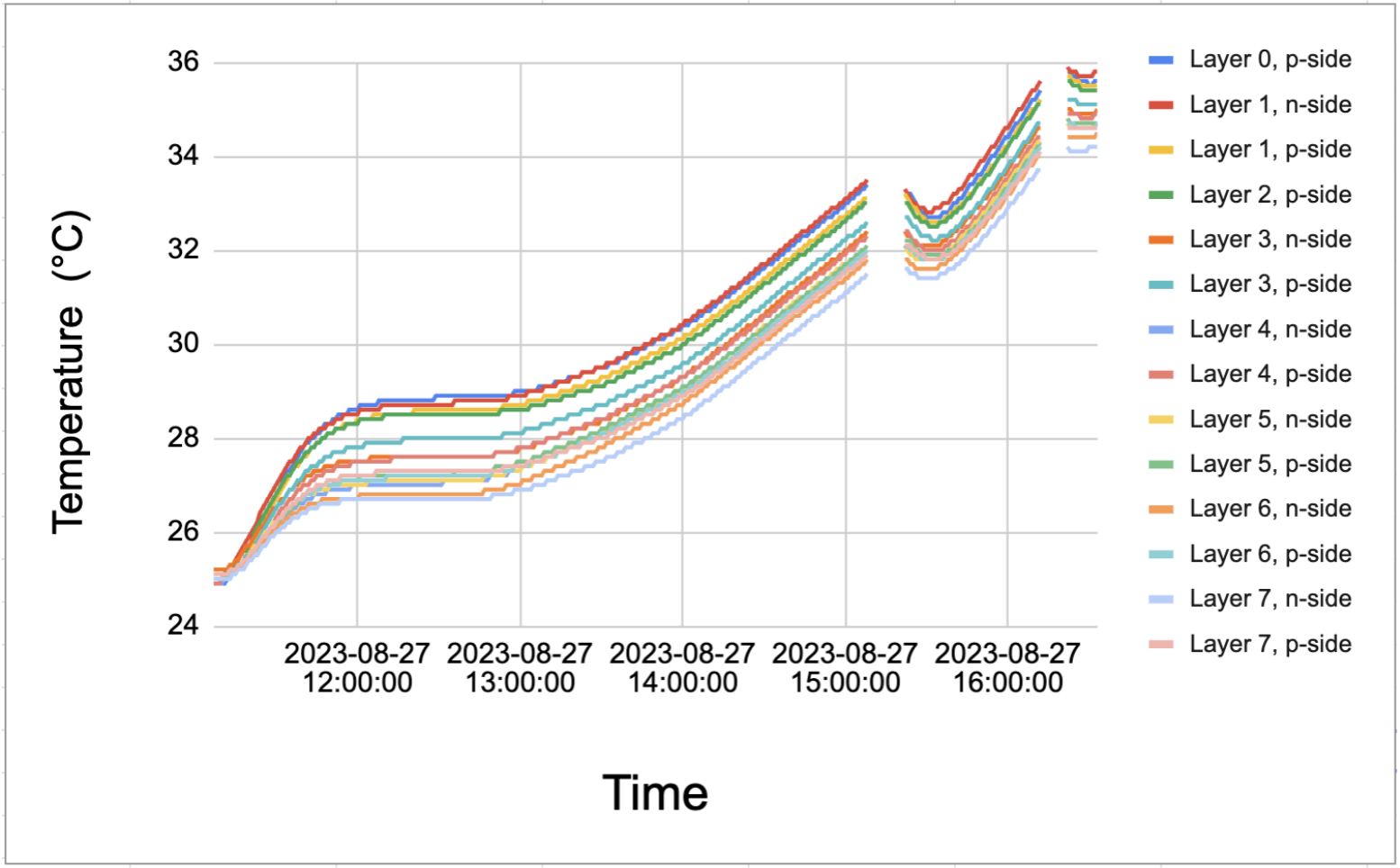}
    \subcaption{Tracker temperature during flight}
    \label{fig:Flight_Temps}
  \end{minipage}
  \hfill
  \begin{minipage}[b]{0.495\textwidth}
    \includegraphics[width=\textwidth]{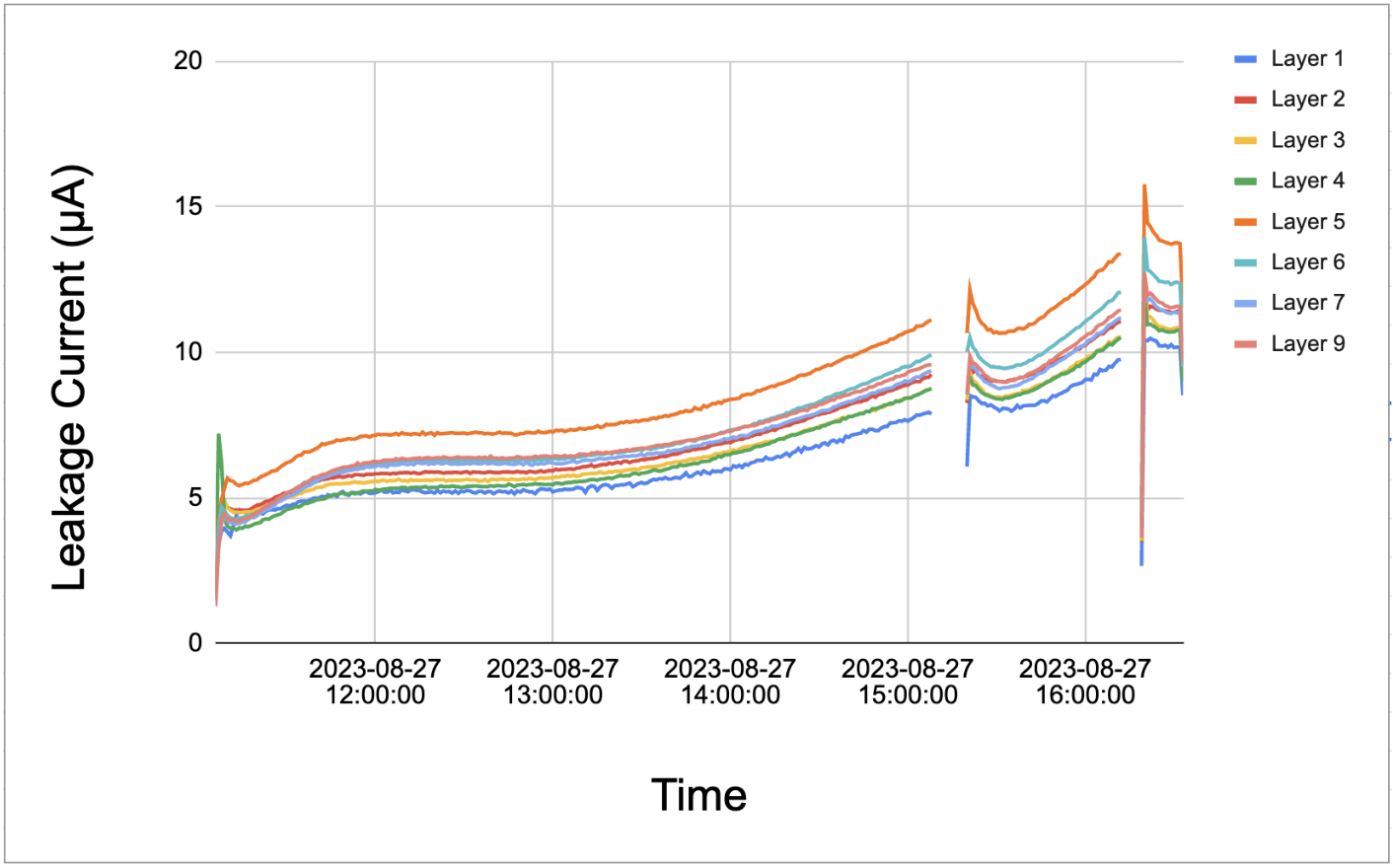}
    \subcaption{Tracker leakage current during flight}
    \label{fig:Flight_Leakage}
  \end{minipage}
  \caption{(a) Temperatures per layer during balloon flight. The system automatically shut off $\sim$4 hours after launch due to the power distribution unit overheating, and then was manually turned on after a 15 minute cool-down period. The system shut off again $\sim$5 hours after launch due to the same issue and was once again powered back on after cooling down. (b) Leakage currents per layer were directly correlated with temperature.}
\end{figure}

The balloon was then brought down in Albuquerque, New Mexico and was subsequently retrieved by NASA's Columbia Scientific Balloon Facility (CSBF). During the flight, $\sim$4 hours after launch, the full {\it ComPair} instrument was powered off due to the main power distribution unit DC-DC converter overheating. After a 15 minute cool-down, the system was started again (Fig.~\ref{fig:Flight_Temps}). This reoccured $\sim$1 hour later as well. Tracker temperatures were also a concern, as each of the layers were operating around the upper end of the functional range by the end of the flight. The Tracker layers' leakage currents were observed to increase during flight as well, presumably due to the increase in temperature (Fig.~\ref{fig:Flight_Leakage}).

\begin{figure}[H]
\centering
    \includegraphics[width=100mm]{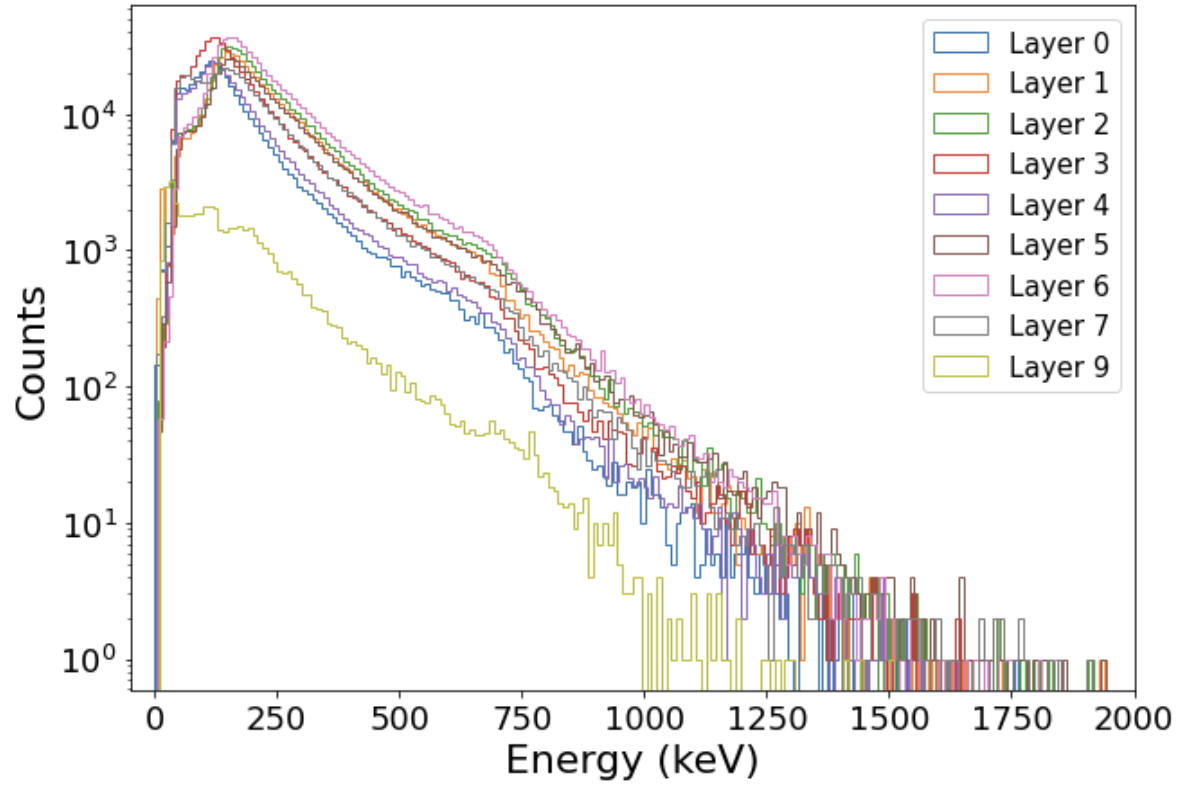}
\caption{Flight spectra from 9 Tracker layers. Each spectra shows the gamma-ray continuum recorded during flight, dominated by albedo gamma-rays. The step-like feature at $\sim$750 keV is the result of an overflow bin energy-cut made on each strip.}
\label{fig:Balloon_Spec}
\end{figure}

Figure \ref{fig:Balloon_Spec} shows the spectra from 9 of the Tracker layers during flight, where individual hit energies are summed for each layer. At float altitude, the background is dominated by albedo gamma-rays, where cosmic rays interact with the surface of Earth and are subsequently reflected back as low energy gamma-rays, so a gamma-ray continuum is expected. It is noted that Layer 9 was the last layer integrated into the Tracker, and therefore the least time was spent troubleshooting noisy strips, which resulted in raised thresholds and lower quality data compared to the other layers. It is also noted that for each strip, the overflow energy bin, corresponding to $\sim$750 keV was cut, hence why there is a small drop off around that energy.

Analysis on the balloon flight data is currently ongoing, but preliminary results look promising. The discrepancy in the spectrum between layers at lower energies ($\lesssim$ 250 keV) is still being explored, but it is believed to be due to different thresholds amongst the layers. 

\section{CONCLUSION}

The {\it ComPair} DSSD Tracker demonstrated its ability to measure low energy gamma-ray events, with adequate position resolution to reconstruct initial photon directions. It then successfully flew as part of the {\it ComPair} instrument, serving as a technology demonstration in a space-like environment, and analysis on the flight data is ongoing. ComPair-2, the next iteration, is currently being developed to serve as a technology demonstration for the MIDEX mission concept AMEGO-X \cite{Caputo_2022}. The Tracker on board ComPair-2 will make use of pixelated silicon detectors as opposed to DSSDs, and will cover a surface area $\sim$4 times larger than its predecessor. ComPair-2 is currently funded up to a beam test that is scheduled to take place in the summer of 2026.

\acknowledgments 
 
This work is supported under NASA Astrophysics Research and Analysis (APRA) grants NNH14ZDA001N-APRA and NNH21ZDA001N-APRA. The material is based upon work supported by NASA under award number 80GSFC21M0002. Research presented in this proceeding was supported by the Laboratory Directed Research and Development program of Los Alamos National Laboratory under project number 20210675ECR. The authors would like to acknowledge NASA Columbia Scientific Balloon Facility for their support during launch, flight, and recovery, as well as the {\it GRAPE} team for accommodating the {\it ComPair} team.

\bibliography{report} 
\bibliographystyle{spiebib} 

\end{document}